\documentclass[a4paper,fleqn]{cas-dc}

\usepackage{graphicx,colordvi}
\usepackage{dcolumn}
\usepackage{bm}
\usepackage{threeparttable}
\usepackage{xspace}
\usepackage{cases}
\usepackage{tabularx,booktabs}
\usepackage{epstopdf}
\usepackage{physics} 
\usepackage{mathtools, amssymb, amsthm, amsmath}
\usepackage{physics}
\usepackage{isotope}

\usepackage[numbers,compress]{natbib}

\def\tsc#1{\csdef{#1}{\textsc{\lowercase{#1}}\xspace}}
\tsc{WGM}
\tsc{QE}

\begin{document}
\let\WriteBookmarks\relax
\def\floatpagepagefraction{1}
\def\textpagefraction{.001}

\shorttitle{Impact of finite-range spin-orbit and tensor terms in Gogny EDF}    

\shortauthors{G. Zietek, N. Pillet, M. Anguiano et~al.}  

\title [mode = title]{Impact of finite-range spin-orbit and tensor terms in Gogny EDF \\ 
on structure and fission properties}

\author[1,2]{G. Zietek}

\author[1,2]{N. Pillet}

\cormark[1]
\ead{nathalie.pillet@cea.fr}

\author[3]{M. Anguiano}

\author[1,2]{P. Carpentier}

\author[1,2]{N. Dubray}

\author[4]{R. N. Bernard}

\author[1,2]{G. Blanchon}

\author[1,2]{D. Regnier}

\affiliation[1]{organization={CEA,DAM,DIF},
            postcode={91297}, 
            city={Arpajon},
            country={France}}

\affiliation[2]{organization={Universit\'e Paris-Saclay, CEA, LMCE},
            postcode={F-91680}, 
            city={Bruy\`eres-le-Ch\^atel},
            country={France}}

\affiliation[3]{organization={Departamento de F\'isica At\'omica, Molecular y Nuclear, Universidad de Granada},
            postcode={18071}, 
            city={Granada},
            country={Spain}}

\affiliation[4]{organization={CEA, DES, IRESNE, DER, SPRC, Cadarache},
            postcode={13108}, 
            city={Saint-Paul-l\`es-Durance},
            country={France}}

\begin{abstract}
Energy Density Functionals are of major interest for the study of the atomic nucleus as, coupled 
with mean-field and beyond N-body approaches, they are applicable to the whole nuclear chart, including superheavy elements. 
On the one hand, the growing need for nuclear data and, on the other hand, the large amount of experimental data on exotic nuclei explain the work carried out on these phenomenological forms of the nucleon-nucleon interaction to analyze the richness of the nuclear phenomena.
In this paper, we propose a fully finite-range extension of the Gogny EDF, including a short-range spin-orbit term and a long-range tensor term. 
The original fitting protocol of the Gogny interaction has been adapted to include both finite range spin-orbit and tensor terms, 
adding new constraints and filters linked to relevant data. Nuclear matter, spectroscopic and fission properties are discussed, highlighting 
ways of improving EDFs when all spin and isospin exchanges are introduced with finite-range terms. 
\end{abstract}


\maketitle

\section{Introduction}\label{intro}

Energy Density Functionals (EDFs) are powerful representations of the nucleon-nucleon effective interaction as they allow to 
capture important nuclear 
short-range correlations at a very low cost, through the introduction of a phenomenological density-dependent central term. 
For example, when guided by {\it g}-matrix 
properties in the (S,T) channels and pertinent nuclear data, they may provide a good starting point for a strategy anchored in mean-field and beyond mean-field
microscopic approaches to solve the N-body nuclear problem. 
The most famous incarnations of EDFs are the Skyrme \cite{SkyrmePhil,Skyrmeoriginal,SkyrmeSO,BrinkVautherin,Vautherin2} and the Gogny \cite{Gogny1,Gogny2,Gogny3,Gogny4,Gogny5} interactions. They have a relatively simple analytical form, including central and spin-orbit terms. 

Since they were proposed and developed essentially starting from the 70's, many works have been accomplished in order to enlarge their domains of 
applicability and improve their precision. Two major directions have been taken, either by generating new parameterizations or by 
extending the analytical expressions. The first direction has been widely followed probably because keeping the analytical form is less 
demanding in terms of global changes and numerical resources. It has brought important improvements in particular with the introduction of the neutron equation of state in the 
fitting procedure, leading to a more realistic description of neutron-rich exotic nuclei and astrophysical properties 
\cite{Chabanat1,Chabanat2,Gogny6,Gogny7,Gogny8,Gogny10,Gogny11,Gogny12,Gogny13,Gogny14}. 
The second direction has shown a renewed interest these last two decades, with the harvest of new data on exotic nuclei, in particular for structure 
and fission studies, that require extended EDFs in order to include more physical properties. 
On the Skyrme side, one can cite the extension of the original second order form \cite{SkyrmePhil,Skyrmeoriginal,SkyrmeSO} to an N$^3$LO order one \cite{Raimondi}. On the Gogny side, one notices the first extension of the D1 original form \cite{Gogny1,Gogny2,Gogny3} with a finite range central 
density-dependent term: the D2 Gogny interaction \cite{Gogny15,Chappertthese,Gogny16} which was accompanied by improvements in nuclear matter (NM) and finite nuclei properties. 
These last two decades, the most striking feature is the large number of works associated with the addition of a tensor term for both types of EDFs.
The first inclusion of a zero-range tensor force in a Skyrme interaction dates back to the 70's and was done by Stancu \textit{et al.}\ \cite{Stancu} on top of the parameterization SIII \cite{SIII}. Afterwards, other attempts were made, either on top of other existing parameterizations \cite{BrownTS, Stancu2, Sly5T, Zalewski1, Zalewski2}, or by carrying out a global refit of the parameters \cite{Lesinski1,Lesinski2,Shen}.
In the study of Onishi and Negele \cite{OnishiNegele}, a tensor force with a Gaussian form factor was added in place of the density-dependent term in a D1-type Gogny interaction. The first addition of a tensor force to the original Gogny interaction D1 was performed by Otsuka \textit{et al.}\ \cite{Gogny17,Gogny18} with a full refit of the parameters. As in the Skyrme EDF case \cite{SIII}, a perturbative inclusion of the tensor term on top of a D1-type Gogny interaction was also tried \cite{Gogny19,Gogny20,Gogny21,Gogny22,Gogny23,Gogny24} by Anguiano \textit{et al.}, including a readjustement of the spin-orbit intensity \cite{MartaMarcella}.
Even if they are less intensively used, one also indicates the M3Y interactions \cite{Bertsch,Anantaraman} in which two spin-orbit and tensor terms with Yukawa form factors are considered. They were only lately applied to mean-field theories, thanks to the work of Nakada \textit{et al.} with the inclusion of a density-dependent term \cite{Nakadaoriginal, NakadaDD, Nakadaplpairing, Nakadafullpairing}. Finally, the covariant EDFs, which are being used more and more, contain also spin-orbit and tensor terms which have a finite-range nature \cite{SHEN2019103713,PhysRevC.106.034315,WANG20242166,Serot,Lalazissis,Nik,Lalazissis2,Nik2,Long,Zhao,PhysRevC.103.054319,PhysRevC.106.L021305}. 

In the original article on the D1 Gogny interaction \cite{Gogny3}, the authors argue on the finite-range choice in building a phenomenological effective interaction, which is supported by the $g$-matrix properties. They start the discussion at the HF level and point out the fact that the long 
range of the force smoothes the fluctuations of the Hartree-Fock (HF) field whereas, in the case of Skyrme forces, a truncation to second order in the density matrix expansion lead to significant differences when compared to the exact density-dependent HF theory \cite{negele}. Even more significant,
the finite-range choice was strongly linked to beyond-HF long-range correlations. Their aim was to define an approach to pairing correlations which was fully self-consistent and general enough to be applicable to the whole nuclear chart, using the Hartree-Fock-Bogoliubov (HFB) formalism.

In the present paper, we report first results obtained with a fully finite-range Gogny interaction including a tensor term, called ``DG'' in the following. Here, the objective of such an extension is to provide a Gogny interaction usable in more general self-consistent theories than HFB which includes all types of nuclear long-range correlations.
After a brief discussion on the fitting procedure, one presents a few standard NM properties. 
Then, highlights on bulk properties in several isotopic chains, the kink in Pb isotopes charge radii, the first excited states in {\it sd}-shell odd nuclei and some fission properties in actinides are proposed to characterize the specific role of the finite range spin-orbit and tensor terms. 
A systematic comparison to D1S and D2 Gogny forces is done. All the mean-field calculations in finite nuclei have been achieved with the HFB3 solver which is an axial HFB code breaking the parity symmetry and which can be used with both one-center (for nuclear structure studies) and two-center (for fission studies) harmonic oscillator bases \cite{dubray2025hfb3axialhfbsolver,dubraycode}. This solver is able to deal with all the various analytical forms of the Gogny EDFs. 
Beyond mean-field and dynamical analyses have employed the multi-configuration self-consistent field approach \cite{pillet1,pillet2,lebloas,robin1,robin2,robin3}, noted MPMH, and the time-dependent generator coordinate method \cite{Gogny4,Gogny5,goutte,regnier,verriere2}, respectively.
Finally, conclusions are drawn and perspectives to this study are suggested. \\

\section{Fitting process}\label{fit}

The fitting protocol used to obtain the DG interaction is a generalization of the original one \cite{Gogny2}, at the origin of the 
parameterizations D1, D1S, D1P, D1N and D2. 
The basic principle consists in sampling a restricted set of parameterizations suited for nuclear physics applications. The algorithm assesses the
quality of a parameterization by computing some of the corresponding properties for nuclear matter and finite nuclei. Mean-field and beyond
properties are targeted while room is left for an explicit treatment of the nuclear long-range correlations  \cite{Gogny2,Gogny15,Chappertthese,Gogny16,Zietektthese}. This will become apparent in the following in the case of binding energies and radii. To reduce the computational burden, the method leverages emulators to estimate some properties in finite nuclei. Constraints and filters are then introduced. At the end, the physicist may pick up one or several parameterizations produced by the sampler, whose ultimate validation is achieved with calculations done for finite nuclei using mean-field and beyond N-body approaches. 

In the original fitting protocol of D1-type Gogny interaction, the constraint sector provides with a set of parameters deduced within the Restricted Hartree-Fock model (HFR), our emulator,
which solves the HF equations assuming pure harmonic oscillator wave functions as single-particle states \cite{Gogny2,Chappertthese,Gogny16}.
Within this approximation, analytical expressions of observables or pseudo-observables are directly expressed in terms of the \{W, B, H, M\} parameter combinations,
defining sets of equations to be inverted to deduce the values of the parameters.
Both binding energies (BE) and charge radii of $^{16}$O and $^{90}$Zr are typically used to provide a first set of equations. 
This makes sense as a linear behavior is found between HF and HFR BE. 
A second set of equations is built from two-body matrix elements (TBME) to control proton and neutron pairing properties and from the energy difference between 
the neutron and the proton 2$s_{1/2}$ levels in $^{48}$Ca which is sensitive to the isospin dependence of the interaction.
Other parameters are left free (\textit{i.e.} the ranges) or chosen by hand (\textit{i.e.} the intensity of the zero-range spin-orbit term).
In the filter sector, many properties coming from infinite or semi-infinite NM are investigated: the saturation density 
$\rho_{0}$, the energy per particle $\mathcal{E}/A$, the incompressibility $K_{\infty}$, the effective masses $m^*/m$, the symmetry energy $a_{\tau}$, 
the total energy by (S,T) channel, the neutron equation of state, the stability criteria and sum rules calculated within the Landau theory of Fermi liquids in view of RPA calculations \cite{padgen}, and the surface energy coefficient for fission applications.

The fitting procedure was adapted for the D2 interaction which introduces a finite-range density-dependent term whose form factor was chosen in order to recover the D1-type zero-range density-dependent term at the limit where the range tends to zero \cite{Chappertthese,Gogny15}. The modifications
included essentially the generalization of the equations for the emulator and the NM properties. The ranges and the parameters \{W, B, H, M\} associated with 
the new density-dependent term were left free, and the intensity of the zero-range spin-orbit term was still chosen by hand.

This D2 fitting procedure has been generalized in turn to the DG analytical form \cite{Zietektthese}, this latter extending the D2 interaction with a finite-range spin-orbit and tensor term. The analytical expression of the DG interaction reads: 

\begin{equation} \label{gognyDG}
\begin{split} 
 v_{12}^{\text{DG}} & \equiv { \sum_{i=1,2} (W_i + B_i P_{\sigma} - H_i  P_{\tau} - M_i P_{\sigma} P_{\tau})} ~v_i(r_{12})\\
 & + (W_3 + B_3 P_{\sigma} - H_3  P_{\tau} - M_3 P_{\sigma} P_{\tau})
~\frac{v_3(r_{12})}{(\mu_3 \sqrt{\pi})^3} ~\rho_{12}  \\
 & + (W_4 - H_4  P_{\tau}) ~\frac{v_4(r_{12})}{- \mu^5_4 \sqrt{\pi}^3/4} ~\vec{L} \cdot \vec{S} \\
 & + (W_5 - H_5  P_{\tau}) ~v_5(r_{12}) ~S_{12} 
\end{split}
\end{equation}
where 
$$v_j(r_{12}) \equiv e^{-(\vec{r}_1 - \vec{r}_2)^2/\mu_j^2}$$
$$ \rho_{12} \equiv \big(\rho^{\alpha}(\vec{r}_1) + \rho^{\alpha}(\vec{r}_2)\big)/2$$
$$ \vec{S} \equiv \big(\vec{\sigma}_1 + \vec{\sigma}_2\big)/2$$
$$\vec{L} \equiv \vec{r}_{12} \times \vec{p}_{12} $$
$$ S_{12} \equiv (\vec{\sigma}_1 \cdot \hat{r}_{12}) (\vec{\sigma}_2 \cdot \hat{r}_{12}) - \vec{\sigma}_1 \cdot \vec{\sigma}_2 /3 $$
and $\vec{p}_{12} = \big(\vec{p}_1-\vec{p}_2\big)/2$, $r_{12}= |\vec{r}_{1}-\vec{r}_{2}|$, $ \hat{r}_{12}= \vec{r}_{12}/r_{12}$.

\noindent The form factor of the spin-orbit term has been chosen in order to recover the zero-range limit used in the original D1-type Gogny interaction when the range $\mu_4$ tends to zero \cite{Zietektthese}.
For the tensor form factor, one has decided to stay in the spirit of the Gogny force with a simple one built from a Gaussian (as it was already done by Anguiano \textit{et al.}), assuming that recovering the one pion exchange potential at large distance has much less impact in finite nuclei 
than in NM \cite{maire}. Moreover, for the expression of S$_{12}$, one has followed the choice of Onishi and Negele \cite{OnishiNegele}.
The DG interaction has a total of twenty-two parameters, among which five ranges and the power $\alpha$ of the nucleon density $\rho$ taken equal to 1/3.
Their values are given in Table. \ref{parameters}.

The fitting procedure of D2 has been modified at two levels for DG \cite{Zietektthese}. The first one has consisted in generalizing the expressions of the D2 fitting code to DG, when
it was relevant. The second one has been dedicated to the determination of the spin-orbit and tensor parameters \{W$_{4}$, H$_{4}$, W$_{5}$, H$_{5}$\}.
In order to do so, a new set of four equations has been created in the fitting protocol implying TBME coupled to T=0 and T=1 channels for various 
total angular momentum values J. This corresponds to a generalization of the equations associated with proton and neutron pairing TBME but for more general
correlations. In particular, it contains TBME related to proton-neutron pairing and quadrupole correlations. 
This has led to the first consistent determination of both spin-orbit and tensor parameters with the other parameters. New filters associated with NM have enriched the fitting process such as the slope of the symmetry energy, the partial wave (PW) decomposition of $\mathcal{E}/A$, 
as well as stability criteria and sum rules, extended by the tensor component in the Landau parameters.
One notes that, in the HF BE used to create the HFR binding energy meta-data, exact Coulomb treatment has been used for DG, unlike D1, D1S and D2, for example, for which the Slater approximation was used. 
Among the various parameterizations outgoing the fitting code,
the final validation of the DG parameterization has been achieved thanks to calculations in finite nuclei.
Several observables have been investigated, among which: (i) at HFB level \cite{dubray2025hfb3axialhfbsolver,dubraycode}: the BE of $^{16}$O, $^{90}$Zr and $^{208}$Pb, 
the absence of BE drift, the absence of pairing energy in doubly magic nuclei, the single-particle energies in $^{40}$Ca, $^{48}$Ca and $^{56}$Ni, the kink in Pb isotope charge radii (ii) at configuration interaction level \cite{pillet1,pillet2,lebloas,robin1,robin2,robin3}: the first excited state energies in even-even and odd $sd$-shell nuclei (iii) at TDGCM+GOA level \cite{Gogny4,Gogny5,goutte,regnier,verriere2}: the height of fission barriers in a few actinides.
The parameter set of the DG interaction is given in TAB.\ref{parameters}. One observes that the fitting code has found a spin-orbit (tensor) range compatible with the omega (pion) meson mass $\mu_{\omega}\simeq$0.25 fm ($\mu_{\pi}\simeq$1.43 fm).

\begin{table}[htb!]
    \begin{tabular}{cccccc}
\hline
        {$i$} & {$\mu_i$}  & {$W_i$} & {$B_i$} & {$H_i$} & {$M_i$} \\
\hline
        1 &  0.80 & -1190.016 & 800.000 & -877.422 & 1198.923 \\
        2 &  1.24 & 109.179 & -191.226 & 133.441 & -277.509  \\
        3 &  0.60 & 1836.200 & 581.600 & 377.600 & -633.220  \\
        4 &  0.20 & 145.483 & {} & 29.634 & {}  \\
        5 &  1.10 & -392.544 & {}  & -196.481 & {}  \\        
\hline
    \end{tabular}
\caption{Values of parameters of the DG Gogny interaction. The index "i" represents the number of the range $\mu_{i}$ (in fm).
For i$\in \{1,2,5\}$, \{$W_i$, $B_i$, $H_i$, $M_i$\} are expressed in MeV, i$=3$ in MeV fm$^3$ and i$=4$ in MeV fm$^5$.}
\label{parameters}
\end{table}

\begin{table}[htb!]
    \begin{tabular}{lccccc}
\hline
                 & {$\mathcal{E}/A$} & {K$_{\infty}$} & m$^*$/$\rm m$ & {a$_{\tau}$} & L\\
                 & {(MeV)} & {(MeV)} & {} & {(MeV)} & {(MeV)}\\
\hline
  DG     &  -16.01 & 210 & 0.74 & 31.3 & 43.02\\
  D2     &  -16.00 & 209 & 0.74 & 31.1 & 44.85 \\
  D1S    &  -16.02 & 210 & 0.70 & 31.1 & 22.43 \\
  D1M    &  -16.03 & 225 & 0.75 & 28.6 & 24.83 \\
  D3G3   &  -16.05 & 227 & 0.68 & 32.6 & 36.70 \\
\hline
  Exp.   &  -16(1) & 215(15) & 0.70(5) & 30(2) & 50(10) \\
\hline
    \end{tabular}
\caption{SNM properties for various Gogny interactions. Experimental data have been extracted from \cite{PQ1,PQ2,PQ3,PQ4,PQ5,MeffMahaux}.}
\label{nmprop}
\end{table}

\section{Nuclear matter properties}\label{nm}

One shows standard properties calculated at the saturation density $\rho_{0}$, namely, the energy 
per particle $\mathcal{E}/A$, the incompressibility $K_{\infty}$, the effective mass $m^*/m$, symmetry energy $a_{\tau}$ and its slope L. 
The values are provided in TAB. \ref{nmprop} for the DG, D2, D1S, D1M and D3G3 parameterizations. Experimental values are also given \cite{PQ1,PQ2,PQ3,PQ4,PQ5,MeffMahaux}. 
All the interactions display similar values for $\rho_{0}$, 0.163 fm$^{-3}$ for DG, D2 and D1S, and 0.165 fm$^{-3}$ for D1M and D3G3.
One sees that the new interaction DG gives satisfactory NM predictions with a value for 
the slope of the symmetry energy compatible with the experimental data. 
The similarity of the central and density-dependent terms of D2 and DG induces that 
many of their NM properties are found very close. 

\begin{figure}[htb!]
\includegraphics[angle=-0,width=26em]{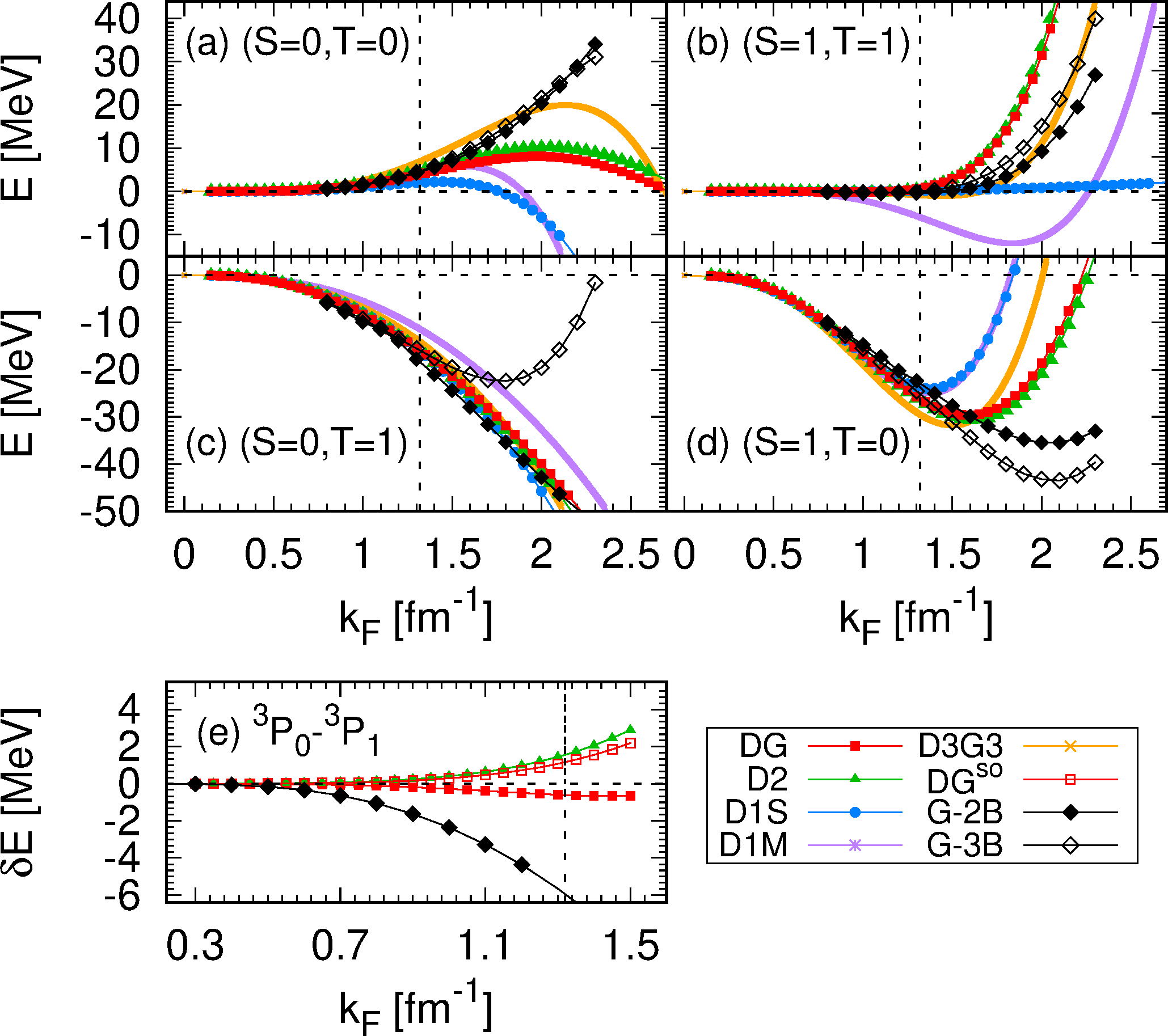}
\caption{\label{fig:epsart} Contributions of spin-isospin channels to the SNM EoS (Panels (a) to (d)). Panel (e)
represents the difference $\delta$E between the  $^3$P$_1$ and $^3$P$_0$ partial waves as defined in the text. Calculations have been done for various Gogny interactions. BBG predictions including 2-body (G-2B) and 3-body (G-3B) contributions are also indicated \cite{Baldo1,vidana,hugo}.}
\label{partial}
\end{figure}

From the (S,T) channel analysis of the symmetric NM (SNM) equation of state (EoS) shown in FIG.\ref{partial}, one sees, in the panels (a) to (d), that the even channels calculated with DG follow in a satisfactory 
way the Bethe-Brueckner-Goldstone (BBG) predictions when 2- and 3-body interactions (noted G-2B and G-3B, respectively) are taken into account respectively
\cite{Baldo1,vidana,hugo}. 
Some small local improvements have been made in the odd ones, starting from D2, in particular for densities higher than the saturation density by postponing the collapse to negative values further in the (S=0,T=0) channel and avoiding the flat (D1S) and too attractive (D1M) behaviors in the (S=1, T=1). However, there is still room for improvement. One observes a correlation between channels which is particularly visible between the odd ones.
In order to go further in the analysis of odd channels in SNM, a PW expansion of $\mathcal{E}/A$ has been carried out \cite{rotival}. 
To highlight the role of the finite-range spin-orbit and tensor terms in odd channels, one has focused on the difference 
$\displaystyle \delta E = \big( \mathcal{E} (^{3}P_0) - \mathcal{E} (^{3}P_1)/3 \big)/A$ between the $^{3}P_0$ and $^{3}P_1$ PW that depends solely on the spin-orbit and tensor contributions \cite{davesne1,davesne2,Zietektthese}. Thus, we have represented solely on FIG.\ref{partial}, panel (e), the DG, D2 and DG without tensor (DG$\rm ^{SO}$) ones. Only DG provides with negative values, in agreement with the 
G-matrix predictions. This effect is explained by the presence of the tensor term whose contribution is attractive in the $^{3}P_0$ and repulsive in the $^{3}P_1$ PWs, leading to a negative contribution to $\delta E$. The contribution of the spin-orbit term is always repulsive in the $^{3}P_0$ 
and $^{3}P_1$ PW, whatever the parameterization, providing with a positive contribution to $\delta E$ whose amplitude is smaller than the tensor one.  

\section{Structure properties}\label{sp}

\begin{figure}[htb!]
\includegraphics[angle=-90,width=26em]{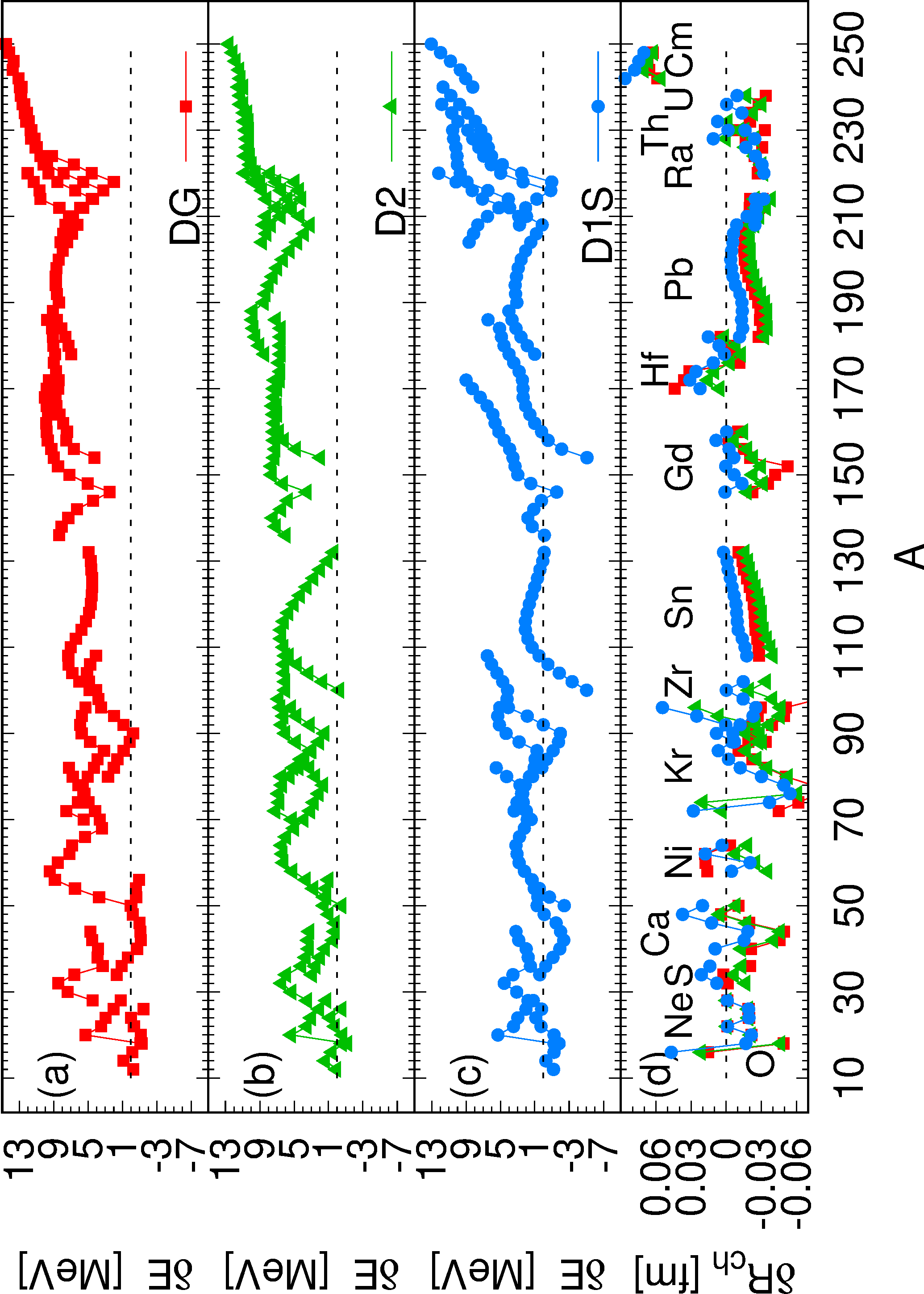}
\caption{Differences between HFB and experimental BE $\delta$E (panels (a)-(c)) and charge radii $\delta R_{\rm ch}$ (panel (d)). Experimental data have been extracted from Refs. \cite{Wang,RadiiExp}.}
\label{fig:BE-R2}
\end{figure}

We discuss now the characteristics of the BE and charge radii for the ground states of several isotopic chains, namely O, S, Ca, Ni, Kr, Zr, Sn, Gd, Hf, Pb, Ra,Th,U and Cm. They have been chosen for their variety in terms of quadrupole deformations. Some are spherical, well-deformed or even show softness with shape coexistence. BE and charge radii have been evaluated at the HFB level. In the fitting process, the constraints on BE and charge radius values are taken smaller than experiment. 
It is desired as nuclear long-range correlations will increase their values in general.
Concerning the BE, the objective was to generate a parameterization which does not contain a drift in energy when compared to experiment, as it was done in the past with the D1N \cite{Gogny7} and D2 \cite{Chappertthese,Gogny15} interactions, allowing a difference of a few MeV with experiment
in order to leave room for correlations. The D1M parameterization went a step further by imposing a $\chi^2$ minimization on 
masses at the 5DCH level \cite{Gogny8}. 
In FIG.\ref{fig:BE-R2}, panels (a) to (c), the difference $\delta {\rm E = E^{HFB}-E^{Exp.}}$ between HFB and experimental BE is represented, respectively for the DG (red squares), D2 (green triangles) and D1S (blue circles) interactions.
Contrary to the D1S interaction, one observes that the drift is fully controlled along isotopic chains for DG and D2. Besides, with the exception of a few light nuclei, the difference is found positive with DG. 
In FIG.\ref{fig:BE-R2}, panel (d), the difference $\delta {\rm R_{ch}= R_{ch}^{HFB}-R_{ch}^{Exp.}}$ between HFB and experimental charge radii
is displayed. 
In the spherical isotopic chains of Ca, Sn and Pb, one clearly sees that the charge radii found with DG and D2 
are smaller than the ones obtained with D1S, revealing a larger difference with experiment. It is desirable as 
charge radii calculated with beyond mean-field calculations have revealed too large values with D1S \cite{delaroche2}. This result is also found in deformed nuclei with 
some exceptions. Among the exceptions, one can cite Kr isotopes which are soft nuclei. They present several minima whose energies are competitive.
In $^{72}$Kr, $^{94}$Kr and $^{96}$Kr, D1S and D2 favors minima with larger quadrupole deformations than DG, leading to strong discrepancies 
with experiment. In the heavy Zr isotopes, which are also soft nuclei, namely $^{98}$Zr, $^{100}$Zr and $^{102}$Zr, the situation is reversed.
DG favors spherical minima leading to too small charge radii whose $\delta {\rm R_{ch}}$ is larger than -0.06 fm. 
Beyond mean-field evaluations will be capital in this area displaying softness.
One remarks also that the predictions exceed the experimental values in light and well-deformed 
$^{170}$Hf, $^{172}$Hf and $^{174}$Hf isotopes.
The effect is even more striking in Cm isotopes, whatever the interaction. The experimental error is larger in 
those Hh and Cm isotopes ($\simeq$2~10$^{-2}$) than in all the other cases ($\simeq$10$^{-3}$).

\begin{figure}[htb!]
\includegraphics[angle=-90,width=26em]{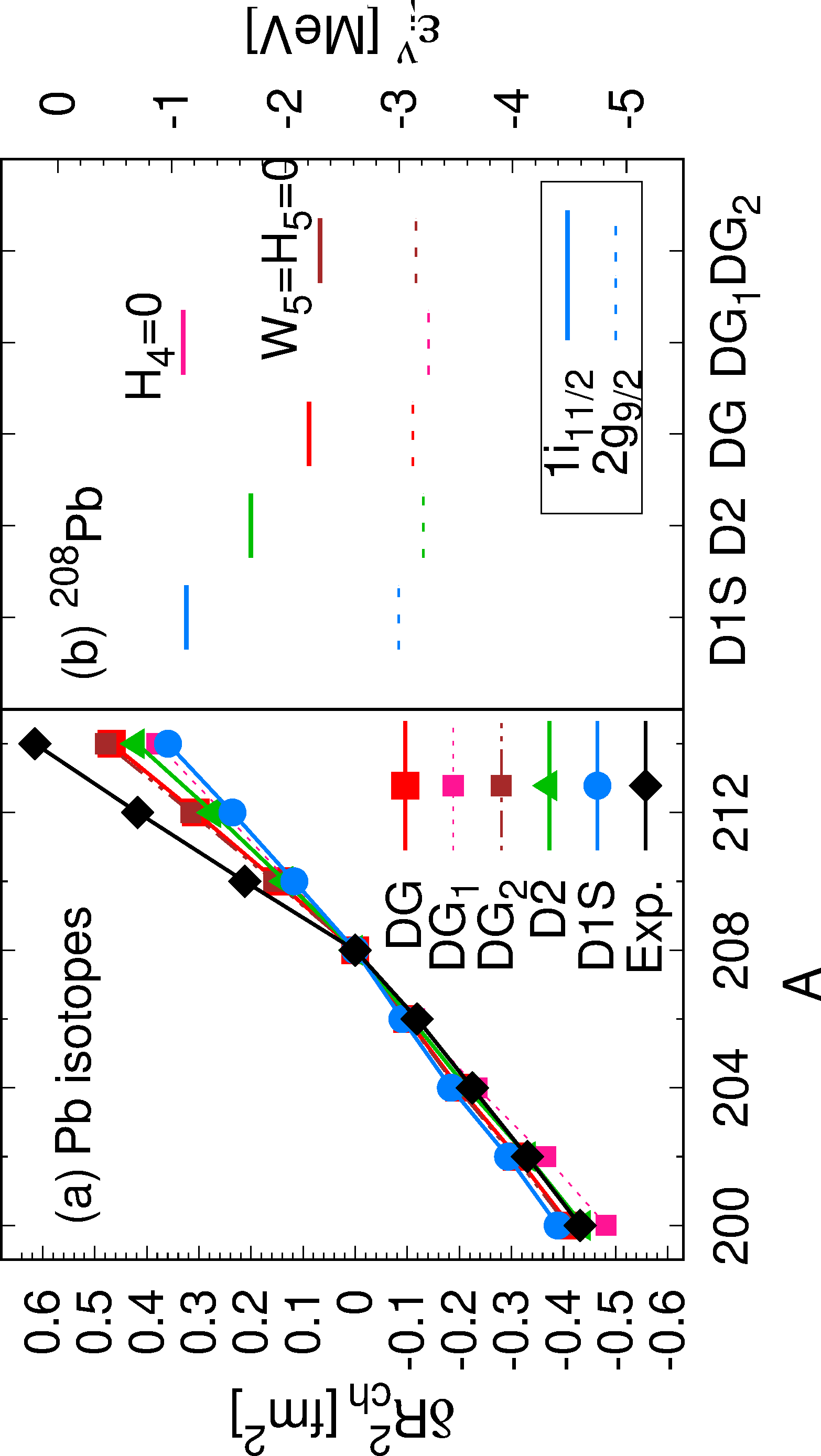}
\caption{\label{fig:kink} Panel (a): Kink $\delta {\rm R^{2}_{ch}}$ in Pb isotopes. Experimental data have been extracted from \cite{RadiiExp}.
Panel (b): The 1i$_{11/2}$-2g$_{9/2}$ gap in $^{208}$Pb. A comparison is done for D1S, D2 and DG Gogny interactions (see text).}
\end{figure}

Detailed structure phenomena are  stringent EDF tests. The kink in Pb isotope charge radii is one of them. It is 
the most widely discussed case in the literature, in particular with the Skyrme and relativistic interactions \cite{Sharma,Sly5T,Lesinski1,Naito,Ebran16}. In our
case, it has been used as an experimental filter to select pertinent sets of the spin-orbit parameters for DG (in particular $H_4$, see \cite{Sharma}) 
and to improve its description with Gogny interactions at the HFB level. 
The formula used to calculate the charge radii can be found in Refs.\cite{Zietektthese,delaroche2,negele1970}. Recent measurements of the mean square radius of the proton \cite{Masspn} and of the neutron \cite{rmsneutron} have been used. 
The difference in charge radii $\delta {\rm R_{ch}^{2} (A)= R_{ch}^{2}(A)- R_{ch}^{2}(208)}$, in which $^{208}$Pb is taken as the reference isotope, is shown
in FIG. \ref{fig:kink}, panel (a), for DG (red squares), D2 (green triangles) and D1S (blue circles), and compared to experiment (black diamonds).
Two additional calculations with DG imposing $H_4=0$, noted DG$_1$ (pink dashed line with squares), $W_5=H_5=0$, noted DG$_2$ (brown dot-dashed line with squares), are also
indicated. 
One clearly sees that DG improves the kink with an angle of $\theta=1.51$ degree, to be compared to $\theta=0.87 $ degree for D2 and $\theta=0.64 degree$ for D1S. Setting $H_4=0$ in DG destroys the improvement, leading to results similar to D1S, whereas setting $W_5=H_5=0$ has essentially no effect. It is well-known that the kink is strongly related to the value of the gap between the 1i$_{11/2}$ and 2g$_{9/2}$ shells, thus implying
a crucial role of pairing correlations that will be more intense with smaller gaps and that will be at the origin of 1i$_{11/2}$ shell occupancy.
For Gogny EDFs, the single-particle levels are given in the right order. They are represented in FIG.\ref{fig:kink}, panel (b), in the case of $^{208}$Pb
for D1S, D2, DG, DG$_1$ and DG$_2$. One observes that their gap is clearly decreased when going from D1S to D2 and DG. 
It is equal to 1.867 MeV, 1.514 MeV and 0.925 MeV, respectively. This gap is also shown for DG with $H_4=0$ and $W_5 = H_5 = 0$.
In a first interpretation, the kink is very sensitive to the spin-orbit intensity in the T=1 channel that acts between the spin-orbit partners
1i$_{11/2}$ and 1i$_{13/2}$ on one side, and 2g$_{7/2}$ and 2g$_{9/2}$ on the other side.
Actually, it has been reduced from $W_{LS}=130$ MeV fm$^5$ for D1S and D2 to $W_4 - H_4 = 115.849$ MeV fm$^5$ for DG. 
This reduction is partly controlled with the isospin exchange parameter $H_4$. It causes a decrease of the gaps between the spin-orbit partners, implying 
a smaller gap between the 1i$_{11/2}$ and 2g$_{9/2}$ shells. One remarks a stronger effect on the 1i$_{11/2}$ shell than the 2g$_{9/2}$ one.
As seen from FIG.\ref{fig:kink}, putting $H_4=0$ leads to a strong increase of the 1i$_{11/2}$-2g$_{9/2}$ gap (2.156 MeV), whereas imposing 
$W_5 = H_5 = 0$ has essentially no effect on the gap and thus on the kink, in agreement with the results observed on $\delta {\rm R_{ch}^{2}}$.
A complementary effect that may play a role in the kink is the value of the neutron effective mass that controls also the density of states around 
the neutron Fermi level. For an isospin asymmetry $\beta = (N-Z)/A \simeq 0.21$ corresponding to $^{208}$Pb, the neutron effective mass $m^*_\nu/m$ is $\simeq0.78$ 
for DG, $\simeq0.77$ for D2 and $\simeq0.73$ for D1S. As D2 and D1S have the same spin-orbit with the same intensity, the improvement obtained with D2 
in comparison with D1S was attributed to the increase of the neutron effective mass, thus to the central terms of the interaction. 
One notes that the neutron effective mass for this range of isospin asymmetry is found to be the closest of the BBG predictions.
The best reproduction with DG is probably a combination of spin-orbit and neutron effective mass effects. 
Besides, as the interaction has been fully refitted, in particular the spin-orbit and tensor terms have been fitted in a consistent way, it is difficult 
to fully disentangle pure spin-orbit and tensor effects. 
In the context of relativistic EDFs whose building is rather different from the Gogny ones, the reproduction of the experimental kink in Pb isotopes is also interpreted in terms of the 1i$_{11/2}$ shell occupancy \cite{Ebran16,marcos,heitz}. In that case, the spin-orbit is also at the origin of the kink but a more complex mechanism shows up, linked to the small Dirac component which takes into account the pseudo-spin symmetry between the 1i$_{11/2}$ and 2g$_{9/2}$ shells, which does not exist in the Skyrme and Gogny EDFs.\\

\begin{figure}[htb!]
\includegraphics[angle=-90,width=26em]{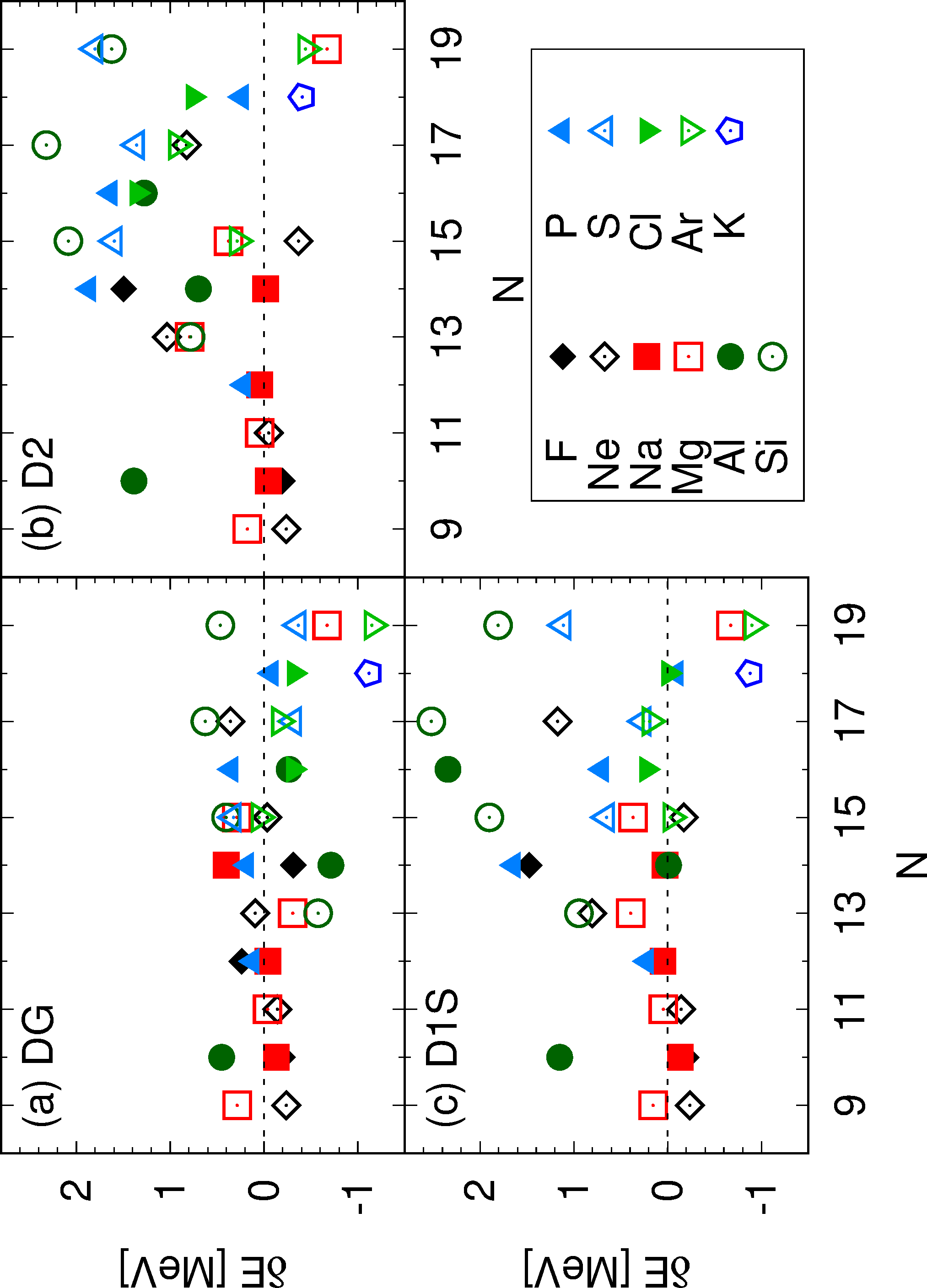}
\caption{\label{fig:oddMPMH} Difference (in MeV) between first experimental excited states and the MPMH ones
predicted with the same total angular momentum and parity. Panels (a) to (c) correspond to the results obtained
with DG, D2 and D1S for odd {\it sd}-shell isotopes.}
\end{figure}

As discussed in the fitting process of the DG interaction, in addition to the previous structure properties, the spin and parity of the ground states as well as the spin, parity and excitation energy of the first excited state in even-even and odd {\it sd}-shell nuclei 
have been used as a filter in the selection of the parameterization \cite{Zietektthese}. These properties have been calculated within a multiconfiguration 
self-consistent field approach \cite{pillet1,pillet2,lebloas,robin1,robin2,robin3}, which is an elaborated Configuration Interaction-type approach allowing in the present case a mixing only in the {\it sd}-shell.
The adding of the excitation energy of the first excited state in the fitting process was motivated by the monopole shift obtained with 
D1S for some even-even isotopes with N and/or Z close 14 and/or 16 \cite{lebloas,robin2}, which was also present with D2. The adding of 
odd isotopes is even a clearier signature of this effect because of the single particle.
Indeed, in many cases, 
their structures have an individual nature (relatively pure particle-hole excitations) at low energy and strongly depend on the value of proton and neutron gaps around the Fermi levels. Thus, the reproduction of the first excited state validates the consistency between mean-field (monopole) and residual (multipole) interaction properties.
In the following, we will focuss on odd nuclei.
Among the sixty odd nuclei pertaining to the {\it sd}-shell, nine of them, namely $\isotope[25]{P}$,
$\isotope[25]{S}$, $\isotope[27]{Cl}$, $\isotope[29]{Cl}$, $\isotope[27]{Ar}$, $\isotope[29]{Ar}$, $\isotope[29]{K}$, $\isotope[31]{K}$ and $\isotope[33]{K}$, are unbound experimentally and have been discarded. Additionally, $\isotope[27]{F}$ and $\isotope[29]{Ne}$ have not been considered either as they display 
negative parity states involving the {\it pf} shell, not taken into account here. The agreement rate obtained for the prediction of the ground state 
angular momenta is $\simeq$81\% for DG, $\simeq$87\% for D1S and $\simeq$89\% for D2, which are similar results. For the first excited states, the agreement 
rates are decreased by $\simeq$20\% for all the interactions, with the appearance of few inversions between the first and second excited states.  
The difference $\Delta \mathcal E$ between the experimental excitation energy of the first excited state and the first state predicted by the MPMH approach sharing the same quantum numbers, are given in FIG.\ref{fig:oddMPMH} for odd proton and neutron nuclei, and calculated with DG, D2 and D1S. A noticeable 
improvement is found for DG. Indeed, the average values $\langle \Delta \mathcal E \rangle$ and the standard deviations 
$\sigma( \Delta \mathcal E )$ are equal to 337 keV and 263 keV for DG respectively, 659 keV and 698 keV for D1S and, 821 keV and 673 keV for D2, which demonstrates the disappearance of the monopole shift. 
This improvement is clearly linked globally to the decrease of the spin-orbit intensity accompanied with a decrease of the 1d$_{5/2}$-1d$_{3/2}$ gap, and locally to the tensor force which has a consequence in particular on the 2s$_{1/2}$-1d$_{3/2}$ gap for some nuclei around N,Z $\simeq$ 14, 16. The same overall trends was found in even-even {\it sd}-shell nuclei with an even better accuracy for the 2$^+_1$ states: $\langle \Delta \mathcal E \rangle \simeq $ 232 keV and $\sigma( \Delta \mathcal E )$= 171 keV for DG, 356 keV and 561 keV for D1S and, 438 keV and 382 keV for D2.
Among the 22 experimentally bound even-even nuclei, the improvement originates from a better description of $^{30}$Si, $^{30}$S and $^{32}$S nuclei for which a monopole shift was obtained in the excited spectra with D1S and D2. This improvement is also linked to the evolution of the gaps, as mentionned in the odd
nuclei case. 
It should be interesting to test the DG-type parameterization with other beyond mean-field approaches which break and restore symmetries 
as it has been done with the D1-type parameterizations \cite{Robledo_2018}.
Besides, a detailed work on the residual (multipole) part of the interaction should be done in the future to improve the accuracy in the prediction of the other excited states.

\section{Fission properties}\label{fp}

Finally, we discuss how the fission properties of actinides behave
with the DG interaction.
We study here the fission barriers and the primary fragment mass yields
obtained within the Gaussian Overlap Approximation of the Time-Dependent Generator Coordinate Method(TDGCM+GOA) \cite{Gogny4,Gogny5,goutte,regnier,verriere2}.
For this purpose, we rely on the standard elongation $\rm Q_{20}$ and asymmetry $\rm Q_{30}$ collective coordinates to describe the fission process.
The fission dynamics results in the evolution of collective wavepackets in a potential landscape given by:
\begin{equation}\label{david1}
  \rm V^{\text{GCM}}(Q_{20},Q_{30}) = E^{\text{HFB}}(Q_{20},Q_{30}) - \epsilon(Q_{20},Q_{30}).
\end{equation}
where $\rm E^{\text{HFB}}$ is the HFB BE and $\epsilon$ is a zero point energy correction.
The first (second) fission barrier height $\rm B_I$ ($\rm B_{II}$) outcoming from this modified potential energy surface reads:
\begin{equation}\label{david2}
\begin{centering}
\rm   B^{GCM}_{I(II)} = V^{GCM}_{B_{I(II)}} - E_0^{GCM}.
\end{centering}
\end{equation}
In Eq.(\ref{david2}), the first term is the potential at the barrier and the second is the GCM+GOA ground state energy~\cite{regnier2}.

\begin{table}[htb!]
  \begin{tabular}{ll|cc|cc}
\hline
    Nucleus & Inter & $\rm B^{HFB}_I$  & $\rm B^{GCM}_I$  & $\rm B^{HFB}_{II}$  & $\rm B^{GCM}_{II}$ \\
    \hline
    $^{236}$U & D1S & 11.0   & 7.7      & 8.0   & 6.5  \\
              & D2  &  9.1   & 7.1      & 7.6   & 6.3   \\
              & DG  &  9.0   & 7.3      & 8.0   & 6.9  \\
\hline
    $^{240}$Pu& D1S & 11.1   & 9.3      & 7.9   & 6.7  \\
              & D2  & 10.5   & 8.6      & 7.6   & 6.4   \\
              & DG  & 10.1   & 8.2      & 8.5   & 7.5  \\
\hline
    $^{252}$Cf& D1S & 11.2   & 9.6      & 3.7   & 2.3   \\
              & D2  & 10.6   & 9.0      & 3.5   & 2.1   \\
              & DG  &  9.3   & 7.6      & 2.8   & 1.9  \\
\hline
  \end{tabular}
  \caption{First and second barrier heights (in MeV) obtained from a pure HFB ($\rm B^{HFB}$) and GCM+GOA ($\rm B^{GCM}$) estimations.}
  \label{tab:energies}
\end{table}

In TAB.\ref{tab:energies}, the values of the GCM barrier heights are reported for $^{236}$U, $^{240}$Pu and $^{252}$Cf. 
For completeness, we compare them to the heights obtained as the difference between the HFB energy at the barrier and the HFB ground state $\rm B_I^{HFB} = E^{HFB}_{B_I} - E_{0}^{HFB}$.
Regardless of the interaction, the GCM first and second barrier heights are systematically lower than the ones estimated at the HFB level 
by $\simeq$1 to 2 MeV.
The introduction of finite-range spin-orbit and tensor terms in DG further tends to decrease the first barrier heights compared to D2, given that these two interactions have very similar central and density-dependent terms. It is difficult to interpret easily this observation that seems to occur in
other actinides.
These results come close to the experimental values, namely 5.0 MeV and 5.7 MeV in $^{236}$U, 6.1 MeV and 5.2 MeV in $^{240}$Pu and 5.3 MeV and 3.5 MeV in $^{252}$Cf \cite{Barriers,GorielyBarrier}.
Indeed, including triaxiality should further improve this consistency by decreasing the first barrier heights (see for example Refs. \cite{fissionchinoise,grams}). In the case of the D1S Gogny interaction, a decrease between to 2 to 4 MeV was obtained \cite{delaroche}.
In the future, it will be interesting to see if the finite-range spin-orbit and tensor terms have an influence on triaxial barrier heights 
when our HFB3 solver will allow to break axiality. 

\begin{figure}[htb!]
\includegraphics[angle=-90,width=26em]{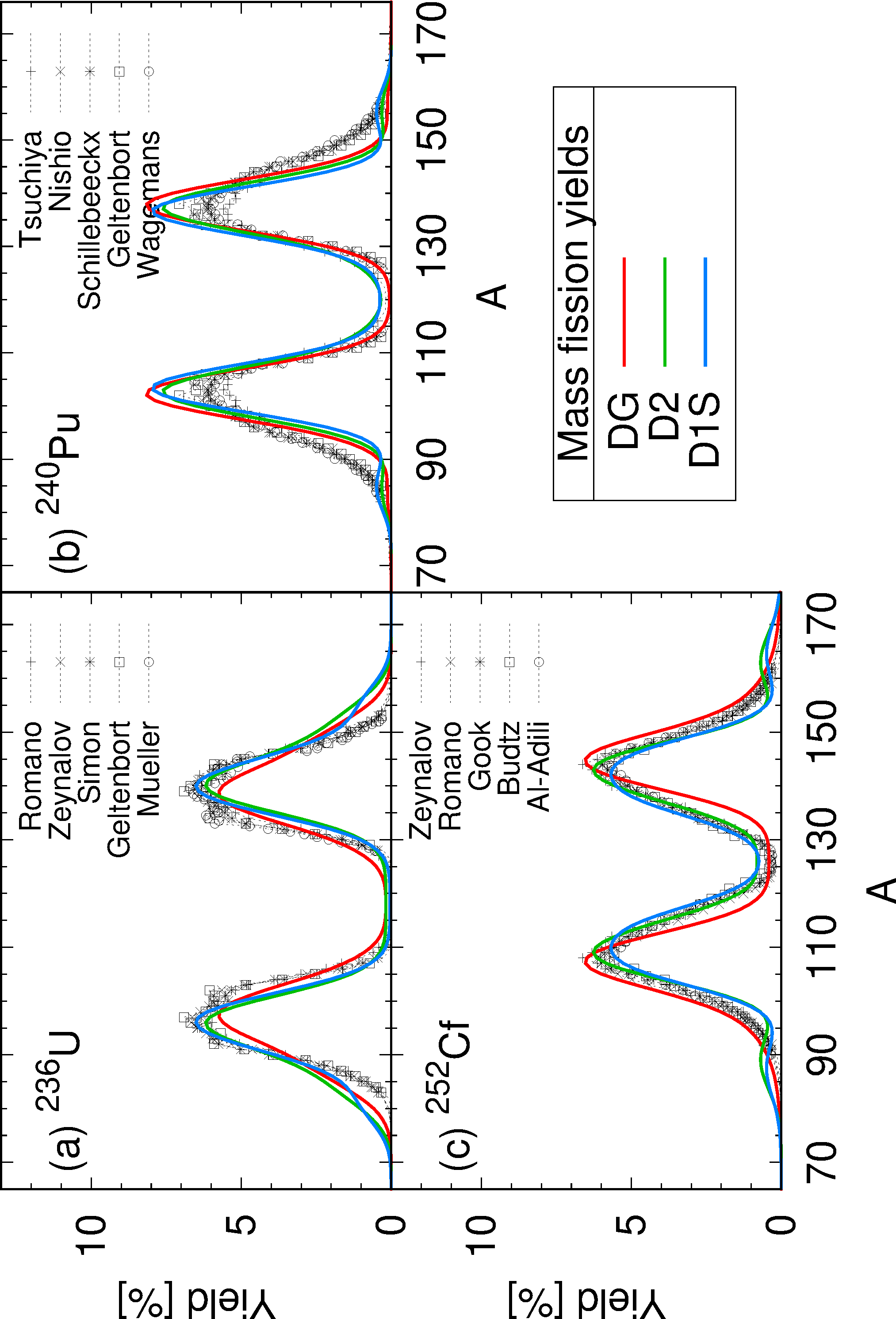}
\caption{\label{fig:yield} Mass fission yields of (a) $^{236}$U, 
(b) $^{240}$Pu and (c) $^{252}$Cf. Calculations have been done for DG, D2 and D1S Gogny interactions.
Experimental data have been extracted from \cite{romanoU,zeynalovU,simon,geltenbortU,mueller,tsuchiya,schillebeeckx,nishio,geltenbort,wagemans,zeylanov,romano,gook,budtz,aladili}.}
\end{figure}

In FIG.\ref{fig:yield}, the fission mass yields of $^{236}$U, $^{240}$Pu and $^{252}$Cf are displayed 
for the three interactions DG, D2 and D1S. They have been evaluated using a scission line characterized by a Q$_{neck}$=5.8 isoline and a 
convolution width of 4 \cite{regnier}. Overall, the results are similar and comparable with experiments 
\cite{romanoU,zeynalovU,simon,geltenbortU,mueller,tsuchiya,schillebeeckx,nishio,geltenbort,wagemans,zeylanov,romano,gook,budtz,aladili}. 
The peaks of the distributions are in relatively good agreement with the experimental ones. \\

\section{Conclusions and perspectives}\label{cp}

In this work, we have presented a new extension of the analytical form of the Gogny interaction, named DG. It consists in a fully finite-range
interaction including a long-range tensor term. In that context, we have generalized the original fitting procedure in order to extract parameterizations. In particular, we have added new constraints and filters in both nuclear matter and finite nuclei. With a selected parameterization, we have analyzed symmetric nuclear matter properties and observables calculated at the mean-field and beyond levels. In particular, we have been interested in masses, charge radii, kink in Pb isotopes, first excited states in odd {\it sd}-shell nuclei and mass fission yields in standard actinides. These studies have allowed to highlight the specific role of the finite-range spin-orbit and tensor terms. In particular, one has obtained improvements in the P partial waves, in the description
of the kink in Pb isotopes, in the excitation energies of low-lying states in even-even and odd $sd$-shell nuclei and a tendency to decrease symmetric fission barrier heights.

For future parameterization optimizations, semi-automatic methods are currently being tested, using penalty functions and bayesian inference. Sensibility studies, using for example Monte Carlo Markov Chains, will also be a part of the selection process to guide the parameterization choice. They will allow for a more exhaustive exploration of the parameters space, the study of uncertainties and  parameter correlations. Furthermore, a study of pairing correlations in the presence of an exact treatment of the Coulomb interaction is also in progress. Finally, it would be interesting to compare to other types of EDF, as Skyrme or
relativistic ones for example.

\paragraph{Aknowledgement} N. Pillet would like to thank JP. Ebran for fruitful discussions on relativistic models.
Computing resources were provided by the CEA, DAM supercomputing facilities.

\printcredits

\bibliographystyle{unsrtnat}

\bibliography{DG_PLB}

@inproceedings{Gogny1,
  title={Proceedings of the International Conference on Nuclear Physics, Munich, August 27-September 1, 1973},
  author = {Gogny, D.},
  editor={de Boer, J. and Mang, H. J. and Maier-Leibnitz, H.},
  pages={48},
  year={1973}
}

@inproceedings{Gogny2,
 title = {Self-consistent pairing calculations},
 author = {Gogny, D.},
 editor = {Ripka, G. and Porneuf, M.},
 year = {1975},
 publisher = {North-Holland Publishing Company},
 country = {Netherlands},
 url = {http://inis.iaea.org/search/search.aspx?orig_q=RN:07230042},
 isbn = {0720403413}
}

@article{Gogny3,
  title = {Hartree-\uppercase{F}ock-\uppercase{B}ogolyubov calculations with the \uppercase{D1} effective interaction on spherical nuclei},
  author = {Decharg\'e, J. and Gogny, D.},
  journal = {Phys. Rev. C},
  volume = {21},
  issue = {4},
  pages = {1568--1593},
  numpages = {0},
  year = {1980},
  month = {Apr},
  publisher = {American Physical Society},
  doi = {10.1103/PhysRevC.21.1568},
  url = {https://link.aps.org/doi/10.1103/PhysRevC.21.1568}
}

@article{Gogny4,
title = {Constrained \uppercase{H}artree-\uppercase{F}ock and beyond},
journal = {Nuclear Physics A},
volume = {502},
pages = {85-104},
year = {1989},
issn = {0375-9474},
doi = {https://doi.org/10.1016/0375-9474(89)90656-8},
url = {https://www.sciencedirect.com/science/article/pii/0375947489906568},
author = {Berger, J.-F. and Girod, M. and Gogny, D.}
}

@article{Gogny5,
title = {Time-dependent quantum collective dynamics applied to nuclear fission},
journal = {Computer Physics Communications},
volume = {63},
number = {1},
pages = {365-374},
year = {1991},
issn = {0010-4655},
doi = {https://doi.org/10.1016/0010-4655(91)90263-K},
url = {https://www.sciencedirect.com/science/article/pii/001046559190263K},
author = {Berger, J.-F. and Girod, M. and Gogny, D.}
}

@article{Gogny6,
doi = {10.1088/0954-3899/25/4/056},
url = {https://dx.doi.org/10.1088/0954-3899/25/4/056},
year = {1999},
month = {apr},
publisher = {},
volume = {25},
number = {4},
pages = {863},
author = {Farine, M. and Von-Eiff, D. and Schuck, P. and Berger, J.-F. and Decharg\'e, J. and Girod, M.},
title = {Towards a new effective interaction of the Gogny type},
journal = {Journal of Physics G: Nuclear and Particle Physics}
}

@article{Gogny7,
    author = "Chappert, F. and Girod, M. and Hilaire, S.",
    title = "{Towards a new \uppercase{G}ogny force parameterization: Impact of the neutron matter equation of state}",
    doi = "10.1016/j.physletb.2008.09.017",
    journal = "Phys. Lett. B",
    volume = "668",
    pages = "420--424",
    year = "2008"
}

@article{Gogny8,
  title = {First \uppercase{G}ogny-\uppercase{H}artree-\uppercase{F}ock-\uppercase{B}ogoliubov Nuclear Mass Model},
  author = {Goriely, S. and Hilaire, S. and Girod, M. and P\'eru, S.},
  journal = {Phys. Rev. Lett.},
  volume = {102},
  issue = {24},
  pages = {242501},
  numpages = {4},
  year = {2009},
  month = {Jun},
  publisher = {American Physical Society},
  doi = {10.1103/PhysRevLett.102.242501},
  url = {https://link.aps.org/doi/10.1103/PhysRevLett.102.242501}
}

@article{Gogny10,
	doi = {10.1016/j.physletb.2018.02.005},
	url = {https://doi.org/10.1016%2Fj.physletb.2018.02.005},
  	year = {2018},
	month = {apr}, 
	publisher = {Elsevier {BV}},
	volume = {779},
	pages = {195-200},
	author = {Gonzalez-Boquera, C. and Centelles, M. and Vi{\~{n}}as, X. and Robledo, L. M.},
	title = {New \uppercase{G}ogny interaction suitable for astrophysical applications},
	journal = {Physics Letters B}
}

@misc{Gogny11,
      title={\uppercase{G}ogny forces in the astrophysical context}, 
      author={Vi\~nas, X. and Gonzalez-Boquera, C. and Centelles, M. and Robledo, L. M. and Mondal, C.},
      year={2018},
      eprint={1810.07469},
      archivePrefix={arXiv},
      primaryClass={nucl-th}
}

@article{Gogny12,
  title = {Structure and composition of the inner crust of neutron stars from \uppercase{G}ogny interactions},
  author = {Mondal, C. and Vi\~nas, X. and Centelles, M. and De, J. N.},
  journal = {Phys. Rev. C},
  volume = {102},
  issue = {1},
  pages = {015802},
  numpages = {18},
  year = {2020},
  month = {Jul},
  publisher = {American Physical Society},
  doi = {10.1103/PhysRevC.102.015802},
  url = {https://link.aps.org/doi/10.1103/PhysRevC.102.015802}
}

@article{Gogny13,
  title = {Higher-order symmetry energy and neutron star core-crust transition with \uppercase{G}ogny forces},
  author = {Gonzalez-Boquera, C. and Centelles, M. and Vi\~nas, X. and Rios, A.},
  journal = {Phys. Rev. C},
  volume = {96},
  issue = {6},
  pages = {065806},
  numpages = {22},
  year = {2017},
  month = {Dec},
  publisher = {American Physical Society},
  doi = {10.1103/PhysRevC.96.065806},
  url = {https://link.aps.org/doi/10.1103/PhysRevC.96.065806}
}

@article{Gogny14,
	url = {https://doi.org/10.1140%2Fepja%2Fs10050-023-01073-w},
	year = {2023},
	month = {jul},
	publisher = {Springer Science and Business Media {LLC}},
	volume = {59},
	number = {7},
	author = {Batail, L. and Davesne, D. and P{\'{e}}ru, S. and Becker, P. and Pastore, A. and Navarro, J.},
	title = {A three-ranged \uppercase{G}ogny interaction in touch with pion exchange: promising results to improve infinite matter properties},
	journal = {The European Physical Journal A}
}

@phdthesis{Chappertthese,
  TITLE = {Nouvelles param{\'e}trisations de l'interaction nucl{\'e}aire effective de \uppercase{G}ogny},
  AUTHOR = {Chappert, F. },
  URL = {https://theses.hal.science/tel-00177379},
  SCHOOL = {{Universit{\'e} Paris Sud - Paris XI}},
  YEAR = {2007},
  MONTH = Jun,
  TYPE = {Theses},
  HAL_ID = {tel-00177379},
  HAL_VERSION = {v1}
}

@article{Gogny15,
  title = {\uppercase{G}ogny force with a finite-range density dependence},
  author = {Chappert, F. and Pillet, N. and Girod, M. and Berger, J.-F.},
  journal = {Phys. Rev. C},
  volume = {91},
  issue = {3},
  pages = {034312},
  numpages = {25},
  year = {2015},
  month = {Mar},
  publisher = {American Physical Society},
  doi = {10.1103/PhysRevC.91.034312},
  url = {https://link.aps.org/doi/10.1103/PhysRevC.91.034312}
}

@article{Gogny16,
  title = {Towards an extended \uppercase{G}ogny force},
  author = {Pillet, N. and Hilaire, S.},
  journal = {The European Physical Journal A},
  volume = {53},
  pages = {193},
  numpages = {10},
  year = {2017},
  month = {Oct},
  doi = {10.1140/epja/i2017-12369-3},
  url = {https://doi.org/10.1140/epja/i2017-12369-3}
}

@article{Gogny17,
  title = {Evolution of Nuclear Shells due to the Tensor Force},
  author = {Otsuka, T. and Suzuki, T. and Fujimoto, R. and Grawe, H. and Akaishi, Y.},
  journal = {Phys. Rev. Lett.},
  volume = {95},
  issue = {23},
  pages = {232502},
  numpages = {4},
  year = {2005},
  month = {Nov},
  publisher = {American Physical Society},
  doi = {10.1103/PhysRevLett.95.232502},
  url = {https://link.aps.org/doi/10.1103/PhysRevLett.95.232502}
}

@article{Gogny18,
title = {Exotic Nuclei and \uppercase{Y}ukawa's Forces},
journal = {Nuclear Physics A},
volume = {805},
number = {1},
pages = {127c-136c},
year = {2008},
note = {INPC 2007},
issn = {0375-9474},
doi = {https://doi.org/10.1016/j.nuclphysa.2008.02.245},
url = {https://www.sciencedirect.com/science/article/pii/S0375947408003345},
author = {Otsuka, T. and Suzuki, T. and Utsuno, Y.}
}

@article{Gogny19,
	url = {https://doi.org/10.1103%2Fphysrevc.83.064306},
	year = {2011},
	month = {jun},
	publisher = {American Physical Society ({APS})},
	volume = {83},
	number = {6},  
	author = {Anguiano, M. and Co', G. and De Donno, V. and Lallena, A. M.},
	title = {Tensor effective interaction in self-consistent random-phase approximation calculations},
	journal = {Physical Review C}
}

@article{Gogny20,
	url = {https://doi.org/10.1103%2Fphysrevc.86.054302},
	year = {2012},
	month = {nov},
	publisher = {American Physical Society ({APS})},
	volume = {86},
	number = {5},
	author = {Anguiano, M. and Grasso, M. and Co', G. and De Donno, V. and Lallena, A. M.},
	title = {Tensor and tensor-isospin terms in the effective \uppercase{G}ogny interaction},
	journal = {Physical Review C}
}

@article{Gogny21,
	url = {https://doi.org/10.1140/epja/i2016-16183-1},
	year = {2016},
	volume = {52},
	number = {7},
	author = {Anguiano, M. and Lallena, A. M. and Co', G. and De Donno, V. and Grasso, M. and Bernard, R. N.},
	title = {\uppercase{G}ogny interactions with tensor terms},
	journal = {The European Physical Journal A}
}

@article{Gogny22,
title = {Interplay between tensor force and deformation in even-even nuclei},
journal = {Nuclear Physics A},
volume = {953},
pages = {32-64},
year = {2016},
issn = {0375-9474},
doi = {https://doi.org/10.1016/j.nuclphysa.2016.03.017},
url = {https://www.sciencedirect.com/science/article/pii/S0375947416300033},
author = {Bernard, R. N. and Anguiano, M.}
}

@article{Gogny23,
	url = {https://doi.org/10.1103%2Fphysrevc.101.044615},
	year = {2020},
	month = {apr},
	publisher = {American Physical Society ({APS})},
	volume = {101},
	number = {4},
	author = {Bernard, R. N. and Pillet, N. and Robledo, L. M. and Anguiano, M.},
	title = {Description of the asymmetric to symmetric fission transition in the neutron-deficient thorium isotopes: Role of the tensor force},
	journal = {Physical Review C}
}

@article{Gogny24,
  title = {Tensor force and deformation in even-even nuclei},
  author = {Co', G. and Anguiano, M. and Lallena, A. M.},
  journal = {Phys. Rev. C},
  volume = {104},
  issue = {1},
  pages = {014313},
  numpages = {18},
  year = {2021},
  month = {Jul},
  publisher = {American Physical Society},
  doi = {10.1103/PhysRevC.104.014313},
  url = {https://link.aps.org/doi/10.1103/PhysRevC.104.014313}
}

@article{OnishiNegele,
title = {Two-body and three-body effective interactions in nuclei},
journal = {Nuclear Physics A},
volume = {301},
number = {2},
pages = {336-348},
year = {1978},
issn = {0375-9474},
doi = {https://doi.org/10.1016/0375-9474(78)90266-X},
url = {https://www.sciencedirect.com/science/article/pii/037594747890266X},
author = {Onishi, N. and Negele, J. W.}
}

@article{negele,
  title = {Density-Matrix Expansion for an Effective Nuclear Hamiltonian},
  author = {Negele, J. W. and Vautherin, D.},
  journal = {Phys. Rev. C},
  volume = {5},
  issue = {5},
  pages = {1472--1493},
  numpages = {0},
  year = {1972},
  month = {May},
  publisher = {American Physical Society},
  doi = {10.1103/PhysRevC.5.1472},
  url = {https://link.aps.org/doi/10.1103/PhysRevC.5.1472}
}

@article{MartaMarcella,
  title = {Tensor parameters in \uppercase{S}kyrme and \uppercase{G}ogny effective interactions: Trends from a ground-state-focused study},
  author = {Grasso, M. and Anguiano, M.},
  journal = {Physical Review C},
  volume = {88},
  issue = {5},
  pages = {054328},
  numpages = {6},
  year = {2013},
  month = {Nov},
  publisher = {American Physical Society},
  doi = {10.1103/PhysRevC.88.054328},
  url = {https://link.aps.org/doi/10.1103/PhysRevC.88.054328}
}

@article{padgen,
    author = {Gogny, D. and Padjen, R.},
    title = {The propagation and damping of the collective modes in nuclear matter},
    doi = {10.1016/0375-9474(77)90104-X},
    journal = {Nucl. Phys. A},
    volume = {293},
    pages = {365--378},
    year = {1977}
}

@phdthesis{maire,
  TITLE = {Etude des propri\'et\'es de l’\'etat fondamental des
noyaux sph\'eriques, par la m\'ethode des perturbations, avec
une interaction nucl\'eon-nucl\'eon r\'ealiste et non singuli\`ere},
  AUTHOR = {Maire, M.},
  SCHOOL = {Universit\'e Paris-Sud},
  YEAR = {1976},
  MONTH = Dec
}

@article{PQ1,
title = {Nuclear masses and deformations},
journal = {Nuclear Physics},
volume = {81},
number = {2},
pages = {1-60},
year = {1966},
issn = {0029-5582},
doi = {https://doi.org/10.1016/S0029-5582(66)80001-9},
url = {https://www.sciencedirect.com/science/article/pii/S0029558266800019},
author = {Myers, W. D. and Swiatecki, W. J.}
}

@article{PQ2,
title = {Average nuclear properties},
journal = {Annals of Physics},
volume = {55},
number = {3},
pages = {395-505},
year = {1969},
issn = {0003-4916},
doi = {https://doi.org/10.1016/0003-4916(69)90202-4},
url = {https://www.sciencedirect.com/science/article/pii/0003491669902024},
author = {Myers, W. D. and Swiatecki, W. J.}
}

@article{PQ3,
title = {Droplet model nuclear density distributions and single-particle potential wells},
journal = {Nuclear Physics A},
volume = {145},
number = {2},
pages = {387-400},
year = {1970},
issn = {0375-9474},
doi = {https://doi.org/10.1016/0375-9474(70)90432-X},
url = {https://www.sciencedirect.com/science/article/pii/037594747090432X},
author = {Myers, W. D.}
}

@article{PQ4,
title = {Global nuclear-structure calculations},
journal = {Nuclear Physics A},
volume = {520},
pages = {c369-c376},
year = {1990},
note = {Nuclear Structure in the Nineties},
issn = {0375-9474},
doi = {https://doi.org/10.1016/0375-9474(90)91161-J},
url = {https://www.sciencedirect.com/science/article/pii/037594749091161J},
author = {M\"oller, P. and Rayford Nix, J.}
}

@article{PQ5,
title = {Nuclear Ground-State Masses and Deformations},
journal = {Atomic Data and Nuclear Data Tables},
volume = {59},
number = {2},
pages = {185-381},
year = {1995},
issn = {0092-640X},
doi = {https://doi.org/10.1006/adnd.1995.1002},
url = {https://www.sciencedirect.com/science/article/pii/S0092640X85710029},
author = {M\"oller, P. and Rayford Nix, J. and Myers, W. D. and Swiatecki, W. J.}
}

@Inbook{MeffMahaux,
author = {Mahaux, C. and Sartor, R.},
editor = {Negele, J. W. and Vogt, E.},
title = {Single-Particle Motion in Nuclei},
bookTitle = {Advances in Nuclear Physics},
year = {1991},
publisher = {Springer US},
address = {Boston, MA},
pages = {1--223},
isbn = {978-1-4613-9910-0},
doi = {10.1007/978-1-4613-9910-0_1},
url = {https://doi.org/10.1007/978-1-4613-9910-0_1}
}

@phdthesis{Zietektthese,
  TITLE = {Towards a generalized effective nuclear \uppercase{G}ogny interaction extended to finite-range spin-orbit and tensor forces},
  AUTHOR = {Zietek, G. },
  URL = {https://theses.hal.science/tel-00177379},
  SCHOOL = {{Universit{\'e} Paris-Saclay}},
  YEAR = {2023},
  MONTH = {Jun},
  TYPE = {Theses},
  HAL_ID = {tel-00177379},
  HAL_VERSION = {v1}
}

@article{Masspn,
  title = {CODATA recommended values of the fundamental physical constants: 2018},
  author = {Tiesinga, E. and Mohr, P. J. and Newell, D. B. and Taylor, B. N.},
  journal = {Reviews of Modern Physics},
  volume = {93},
  issue = {2},
  pages = {025010},
  numpages = {63},
  year = {2021},
  month = {Jun},
  publisher = {American Physical Society},
  doi = {10.1103/RevModPhys.93.025010},
  url = {https://link.aps.org/doi/10.1103/RevModPhys.93.025010}
}

@article{rmsneutron,
title = {Measurement of the neutron charge radius and the role of its constituents},
author = {Atac, H. and Constantinou, M. and Meziani, Z.-E. and Paolone, M. and Sparveris, N.},
journal = {Nature Communications},
volume = {12},
number = {1759},
issue = {1},
year = {2021},
issn = {2041-1723},
url = {https://doi.org/10.1038/s41467-021-22028-z}
}

@article{negele1970,
  title = {Structure of Finite Nuclei in the Local-Density Approximation},
  author = {Negele, J. W.},
  journal = {Phys. Rev. C},
  volume = {1},
  issue = {4},
  pages = {1260--1321},
  numpages = {0},
  year = {1970},
  month = {Apr},
  publisher = {American Physical Society},
  doi = {10.1103/PhysRevC.1.1260},
  url = {https://link.aps.org/doi/10.1103/PhysRevC.1.1260}
}

@article{marcos,
  TITLE = {Relativistic effects on the kink of nuclear charge radii in lead},
  AUTHOR = {Marcos, S. and Niembro, R. and L\'opez-Quelle, M.},
  JOURNAL = {{Eur.Phys.J.A}},
  VOLUME = {60},
  PAGES = {78},
  YEAR = {2024},
  DOI = {10.1140/epja/s10050-024-01304-8},
  url ={https://doi.org/10.1140/epja/s10050-024-01304-8}
}

@phdthesis{rotival,
  TITLE = {{Fonctionnelles d'\'energie non-empiriques pour la structure nucléaire}},
  AUTHOR = {Rotival, V.},
  URL = {https://theses.hal.science/tel-00409482},
  SCHOOL = {{Universit{\'e} Paris-Diderot - Paris VII}},
  YEAR = {2008},
  MONTH = Sep,
  TYPE = {Theses},
  HAL_ID = {tel-00409482},
  HAL_VERSION = {v1},
}

@article{davesne1,
doi = {10.1088/0031-8949/90/11/114002},
url = {https://dx.doi.org/10.1088/0031-8949/90/11/114002},
year = {2015},
month = {oct},
publisher = {IOP Publishing},
volume = {90},
number = {11},
pages = {114002},
author = {Davesne, D. and Meyer, J. and Pastore, A. and Navarro, J.},
title = {Partial wave decomposition of the N3LO equation of state},
journal = {Physica Scripta}
}

@article{davesne2,
title = {Infinite matter properties and zero-range limit of non-relativistic finite-range interactions},
journal = {Annals of Physics},
volume = {375},
pages = {288-312},
year = {2016},
issn = {0003-4916},
doi = {https://doi.org/10.1016/j.aop.2016.10.013},
url = {https://www.sciencedirect.com/science/article/pii/S0003491616302305},
author = {Davesne, D. and Becker, P. and Pastore, A. and Navarro, J.}
}

@article{SkyrmePhil,
	author = {T. H. R. Skyrme},
	doi = {10.1080/14786435608238186},
	journal = {Philosophical Magazine},
	number = {11},
	pages = {1043--1054},
	title = {Cvii. The Nuclear Surface},
	volume = {1},
	year = {1956}
}

@article{Skyrmeoriginal,
title = {The effective nuclear potential},
journal = {Nuclear Physics},
volume = {9},
number = {4},
pages = {615-634},
year = {1958},
issn = {0029-5582},
doi = {https://doi.org/10.1016/0029-5582(58)90345-6},
url = {https://www.sciencedirect.com/science/article/pii/0029558258903456},
author = {T. H. R. Skyrme}
}

@article{SkyrmeSO,
title = {The spin-orbit interaction in nuclei},
journal = {Nuclear Physics},
volume = {9},
number = {4},
pages = {635-640},
year = {1958},
issn = {0029-5582},
doi = {https://doi.org/10.1016/0029-5582(58)90346-8},
url = {https://www.sciencedirect.com/science/article/pii/0029558258903468},
author = {T. H. R. Skyrme}
}

@article{BrinkVautherin,
  title = {Hartree-\uppercase{F}ock Calculations with \uppercase{S}kyrme's Interaction. I. Spherical Nuclei},
  author = {Vautherin, D. and Brink, D. M.},
  journal = {Phys. Rev. C},
  volume = {5},
  issue = {3},
  pages = {626--647},
  numpages = {0},
  year = {1972},
  month = {Mar},
  publisher = {American Physical Society},
  doi = {10.1103/PhysRevC.5.626},
  url = {https://link.aps.org/doi/10.1103/PhysRevC.5.626}
}

@article{Vautherin2,
  title = {Hartree-\uppercase{F}ock Calculations with \uppercase{S}kyrme's Interaction. II. Axially Deformed Nuclei},
  author = {Vautherin, D.},
  journal = {Phys. Rev. C},
  volume = {7},
  issue = {1},
  pages = {296--316},
  numpages = {0},
  year = {1973},
  month = {Jan},
  publisher = {American Physical Society},
  doi = {10.1103/PhysRevC.7.296},
  url = {https://link.aps.org/doi/10.1103/PhysRevC.7.296}
}

@article{Chabanat1,
title = {A \uppercase{S}kyrme parametrization from subnuclear to neutron star densities},
journal = {Nuclear Physics A},
volume = {627},
number = {4},
pages = {710-746},
year = {1997},
issn = {0375-9474},
doi = {https://doi.org/10.1016/S0375-9474(97)00596-4},
url = {https://www.sciencedirect.com/science/article/pii/S0375947497005964},
author = {E. Chabanat and P. Bonche and P. Haensel and J. Meyer and R. Schaeffer},
}

@article{Chabanat2,
title = {A \uppercase{S}kyrme parametrization from subnuclear to neutron star densities Part II. Nuclei far from stabilities},
journal = {Nuclear Physics A},
volume = {635},
number = {1},
pages = {231-256},
year = {1998},
issn = {0375-9474},
doi = {https://doi.org/10.1016/S0375-9474(98)00180-8},
url = {https://www.sciencedirect.com/science/article/pii/S0375947498001808},
author = {E. Chabanat and P. Bonche and P. Haensel and J. Meyer and R. Schaeffer},
}

@article{Lesinski1,
  title = {Tensor part of the \uppercase{S}kyrme energy density functional: Spherical nuclei},
  author = {Lesinski, T. and Bender, M. and Bennaceur, K. and Duguet, T. and Meyer, J.},
  journal = {Phys. Rev. C},
  volume = {76},
  issue = {1},
  pages = {014312},
  numpages = {34},
  year = {2007},
  month = {Jul},
  publisher = {American Physical Society},
  doi = {10.1103/PhysRevC.76.014312},
  url = {https://link.aps.org/doi/10.1103/PhysRevC.76.014312}
}

@article{Lesinski2,
  title = {Tensor part of the \uppercase{S}kyrme energy density functional. II. Deformation properties of magic and semi-magic nuclei},
  author = {Bender, M. and Bennaceur, K. and Duguet, T. and Heenen, P. -H. and Lesinski, T. and Meyer, J.},
  journal = {Phys. Rev. C},
  volume = {80},
  issue = {6},
  pages = {064302},
  numpages = {30},
  year = {2009},
  month = {Dec},
  publisher = {American Physical Society},
  doi = {10.1103/PhysRevC.80.064302},
  url = {https://link.aps.org/doi/10.1103/PhysRevC.80.064302}
}

@article{Raimondi,
  title = {Effective pseudopotential for energy density functionals with higher-order derivatives},
  author = {Raimondi, F. and Carlsson, B. G. and Dobaczewski, J.},
  journal = {Phys. Rev. C},
  volume = {83},
  issue = {5},
  pages = {054311},
  numpages = {15},
  year = {2011},
  month = {May},
  publisher = {American Physical Society},
  doi = {10.1103/PhysRevC.83.054311},
  url = {https://link.aps.org/doi/10.1103/PhysRevC.83.054311}
}

@article{Shen,
  title = {\uppercase{S}kyrme functional with tensor terms from ab initio calculations of neutron-proton drops},
  author = {Shen, Shihang and Col\`o, Gianluca and Roca-Maza, Xavier},
  journal = {Phys. Rev. C},
  volume = {99},
  issue = {3},
  pages = {034322},
  numpages = {15},
  year = {2019},
  month = {Mar},
  publisher = {American Physical Society},
  doi = {10.1103/PhysRevC.99.034322},
  url = {https://link.aps.org/doi/10.1103/PhysRevC.99.034322}
}

@article{Stancu,
title = {The tensor part of \uppercase{S}kyrme's interaction},
journal = {Physics Letters B},
volume = {68},
number = {2},
pages = {108-112},
year = {1977},
issn = {0370-2693},
doi = {https://doi.org/10.1016/0370-2693(77)90178-2},
url = {https://www.sciencedirect.com/science/article/pii/0370269377901782},
author = {Stancu, Fl. and Brink, D. M. and Flocard, H.}
}

@article{SIII,
title = {Nuclear ground-state properties and self-consistent calculations with the \uppercase{S}kyrme interaction: (I). Spherical description},
journal = {Nuclear Physics A},
volume = {238},
number = {1},
pages = {29-69},
year = {1975},
issn = {0375-9474},
doi = {https://doi.org/10.1016/0375-9474(75)90338-3},
url = {https://www.sciencedirect.com/science/article/pii/0375947475903383},
author = {Beiner, M. and Flocard, H. and {Van Giai}, N. and Quentin, P.}
}

@article{BrownTS,
  title = {Tensor interaction contributions to single-particle energies},
  author = {Brown, B. A. and Duguet, T. and Otsuka, T. and Abe, D. and Suzuki, T.},
  journal = {Physical Review C},
  volume = {74},
  issue = {6},
  pages = {061303},
  numpages = {5},
  year = {2006},
  month = {Dec},
  publisher = {American Physical Society},
  doi = {10.1103/PhysRevC.74.061303},
  url = {https://link.aps.org/doi/10.1103/PhysRevC.74.061303}
}

@article{Stancu2,
  title = {Evolution of nuclear shells with the \uppercase{S}kyrme density dependent interaction},
  author = {Brink, D. M. and Stancu, Fl.},
  journal = {Physical Review C},
  volume = {75},
  issue = {6},
  pages = {064311},
  numpages = {6},
  year = {2007},
  month = {Jun},
  publisher = {American Physical Society},
  doi = {10.1103/PhysRevC.75.064311},
  url = {https://link.aps.org/doi/10.1103/PhysRevC.75.064311}
}

@article{Sly5T,
title = {Spin-orbit splitting and the tensor component of the \uppercase{S}kyrme interaction},
journal = {Physics Letters B},
volume = {646},
number = {5},
pages = {227-231},
year = {2007},
issn = {0370-2693},
doi = {https://doi.org/10.1016/j.physletb.2007.01.033},
url = {https://www.sciencedirect.com/science/article/pii/S0370269307001384},
author = {Col\`o, G. and Sagawa, H. and Fracasso, S. and Bortignon, P. F.}
}

@article{Zalewski1,
  title = {Spin-orbit and tensor mean-field effects on spin-orbit splitting including self-consistent core polarizations},
  author = {Zalewski, M. and Dobaczewski, J. and Satu\l{}a, W. and Werner, T. R.},
  journal = {Physical Review C},
  volume = {77},
  issue = {2},
  pages = {024316},
  numpages = {13},
  year = {2008},
  month = {Feb},
  publisher = {American Physical Society},
  doi = {10.1103/PhysRevC.77.024316},
  url = {https://link.aps.org/doi/10.1103/PhysRevC.77.024316}
}

@article{Zalewski2,
  title = {Global nuclear structure effects of the tensor interaction},
  author = {Zalewski, M. and Olbratowski, P. and Rafalski, M. and Satu\l{}a, W. and Werner, T. R. and Wyss, R. A.},
  journal = {Physical Review C},
  volume = {80},
  issue = {6},
  pages = {064307},
  numpages = {10},
  year = {2009},
  month = {Dec},
  publisher = {American Physical Society},
  doi = {10.1103/PhysRevC.80.064307},
  url = {https://link.aps.org/doi/10.1103/PhysRevC.80.064307}
}

@article{Bertsch,
title = {Interactions for inelastic scattering derived from realistic potentials},
journal = {Nuclear Physics A},
volume = {284},
number = {3},
pages = {399-419},
year = {1977},
issn = {0375-9474},
doi = {https://doi.org/10.1016/0375-9474(77)90392-X},
url = {https://www.sciencedirect.com/science/article/pii/037594747790392X},
author = {Bertsch, G. and Borysowicz, J. and McManus, H. and Love, W. G.}
}

@article{Anantaraman,
title = {An effective interaction for inelastic scattering derived from the Paris potential},
journal = {Nuclear Physics A},
volume = {398},
number = {2},
pages = {269-278},
year = {1983},
issn = {0375-9474},
doi = {https://doi.org/10.1016/0375-9474(83)90487-6},
url = {https://www.sciencedirect.com/science/article/pii/0375947483904876},
author = {Anantaraman, N. and Toki, H. and Bertsch, G. F.}
}

@article{Nakadaoriginal,
title = {Hartree-\uppercase{F}ock calculations on unstable nuclei with several types of effective interactions},
journal = {Nuclear Physics A},
volume = {722},
pages = {C117-C122},
year = {2003},
issn = {0375-9474},
doi = {https://doi.org/10.1016/S0375-9474(03)01346-0},
url = {https://www.sciencedirect.com/science/article/pii/S0375947403013460},
author = {Nakada, H.}
}

@article{NakadaDD,
  title = {Hartree-\uppercase{F}ock approach to nuclear matter and finite nuclei with M3Y-type nucleon-nucleon interactions},
  author = {Nakada, H.},
  journal = {Phys. Rev. C},
  volume = {68},
  issue = {1},
  pages = {014316},
  numpages = {14},
  year = {2003},
  month = {Jul},
  publisher = {American Physical Society},
  doi = {10.1103/PhysRevC.68.014316},
  url = {https://link.aps.org/doi/10.1103/PhysRevC.68.014316}
}

@article{Nakadaplpairing,
  title = {Mean-field approach to nuclear structure with semi-realistic nucleon-nucleon interactions},
  author = {Nakada, H.},
  journal = {Phys. Rev. C},
  volume = {78},
  issue = {5},
  pages = {054301},
  numpages = {13},
  year = {2008},
  month = {Nov},
  publisher = {American Physical Society},
  doi = {10.1103/PhysRevC.78.054301},
  url = {https://link.aps.org/doi/10.1103/PhysRevC.78.054301}
}

@article{Nakadafullpairing,
  title = {Modified parameter sets of M3Y-type semi-realistic nucleon-nucleon interactions for nuclear structure studies},
  author = {Nakada, H.},
  journal = {Phys. Rev. C},
  volume = {81},
  issue = {2},
  pages = {027301},
  numpages = {4},
  year = {2010},
  month = {Feb},
  publisher = {American Physical Society},
  doi = {10.1103/PhysRevC.81.027301},
  url = {https://link.aps.org/doi/10.1103/PhysRevC.81.027301}
}

@article{Serot,
    author = "Serot, Brian D. and Walecka, John Dirk",
    title = "{The Relativistic Nuclear Many Body Problem}",
    reportNumber = "ITP-740-STANFORD",
    journal = "Adv. Nucl. Phys.",
    volume = "16",
    pages = "1--327",
    year = "1986"
}

@article{Lalazissis,
  title = {New parametrization for the Lagrangian density of relativistic mean field theory},
  author = {Lalazissis, G. A. and K\"onig, J. and Ring, P.},
  journal = {Phys. Rev. C},
  volume = {55},
  issue = {1},
  pages = {540--543},
  numpages = {0},
  year = {1997},
  month = {Jan},
  publisher = {American Physical Society},
  doi = {10.1103/PhysRevC.55.540},
  url = {https://link.aps.org/doi/10.1103/PhysRevC.55.540}
}

@article{Nik,
  title = {Relativistic \uppercase{H}artree-\uppercase{B}ogoliubov model with density-dependent meson-nucleon couplings},
  author = {Nik\ifmmode \check{s}\else \v{s}\fi{}i\ifmmode \acute{c}\else \'{c}\fi{}, T. and Vretenar, D. and Finelli, P. and Ring, P.},
  journal = {Phys. Rev. C},
  volume = {66},
  issue = {2},
  pages = {024306},
  numpages = {15},
  year = {2002},
  month = {Aug},
  publisher = {American Physical Society},
  doi = {10.1103/PhysRevC.66.024306},
  url = {https://link.aps.org/doi/10.1103/PhysRevC.66.024306}
}

@article{Lalazissis2,
  title = {New relativistic mean-field interaction with density-dependent meson-nucleon couplings},
  author = {Lalazissis, G. A. and Nik\ifmmode \check{s}\else \v{s}\fi{}i\ifmmode \acute{c}\else \'{c}\fi{}, T. and Vretenar, D. and Ring, P.},
  journal = {Phys. Rev. C},
  volume = {71},
  issue = {2},
  pages = {024312},
  numpages = {10},
  year = {2005},
  month = {Feb},
  publisher = {American Physical Society},
  doi = {10.1103/PhysRevC.71.024312},
  url = {https://link.aps.org/doi/10.1103/PhysRevC.71.024312}
}

@article{Nik2,
  title = {Relativistic nuclear energy density functionals: Adjusting parameters to binding energies},
  author = {Nik\ifmmode \check{s}\else \v{s}\fi{}i\ifmmode \acute{c}\else \'{c}\fi{}, T. and Vretenar, D. and Ring, P.},
  journal = {Phys. Rev. C},
  volume = {78},
  issue = {3},
  pages = {034318},
  numpages = {19},
  year = {2008},
  month = {Sep},
  publisher = {American Physical Society},
  doi = {10.1103/PhysRevC.78.034318},
  url = {https://link.aps.org/doi/10.1103/PhysRevC.78.034318}
}

@article{Long,
title = {Density-dependent relativistic \uppercase{H}artree–\uppercase{F}ock approach},
journal = {Physics Letters B},
volume = {640},
number = {4},
pages = {150-154},
year = {2006},
issn = {0370-2693},
doi = {https://doi.org/10.1016/j.physletb.2006.07.064},
url = {https://www.sciencedirect.com/science/article/pii/S0370269306009610},
author = {Wen-Hui Long and Nguyen {Van Giai} and Jie Meng},
}

@article{Zhao,
  title = {New parametrization for the nuclear covariant energy density functional with a point-coupling interaction},
  author = {Zhao, P. W. and Li, Z. P. and Yao, J. M. and Meng, J.},
  journal = {Phys. Rev. C},
  volume = {82},
  issue = {5},
  pages = {054319},
  numpages = {14},
  year = {2010},
  month = {Nov},
  publisher = {American Physical Society},
  doi = {10.1103/PhysRevC.82.054319},
  url = {https://link.aps.org/doi/10.1103/PhysRevC.82.054319}
}

@article{SHEN2019103713,
title = {Towards an ab initio covariant density functional theory for nuclear structure},
journal = {Progress in Particle and Nuclear Physics},
volume = {109},
pages = {103713},
year = {2019},
issn = {0146-6410},
doi = {https://doi.org/10.1016/j.ppnp.2019.103713},
url = {https://www.sciencedirect.com/science/article/pii/S0146641019300481},
author = {Shihang Shen and Haozhao Liang and Wen Hui Long and Jie Meng and Peter Ring}
}

@article{PhysRevC.106.034315,
  title = {Covariant density functional theory with localized exchange terms},
  author = {Zhao, Qiang and Ren, Zhengxue and Zhao, Pengwei and Meng, Jie},
  journal = {Phys. Rev. C},
  volume = {106},
  issue = {3},
  pages = {034315},
  numpages = {15},
  year = {2022},
  month = {Sep},
  publisher = {American Physical Society},
  doi = {10.1103/PhysRevC.106.034315},
  url = {https://link.aps.org/doi/10.1103/PhysRevC.106.034315}
}

@article{WANG20242166,
title = {Tensor-force effects on nuclear matter in relativistic ab initio theory},
journal = {Science Bulletin},
volume = {69},
number = {14},
pages = {2166-2169},
year = {2024},
issn = {2095-9273},
doi = {https://doi.org/10.1016/j.scib.2024.05.013},
url = {https://www.sciencedirect.com/science/article/pii/S2095927324003475},
author = {Sibo Wang and Hui Tong and Chencan Wang and Qiang Zhao and Peter Ring and Jie Meng}
}

@article{PhysRevC.103.054319,
  title = {Nuclear matter in relativistic Brueckner-Hartree-Fock theory with Bonn potential in the full Dirac space},
  author = {Wang, Sibo and Zhao, Qiang and Ring, Peter and Meng, Jie},
  journal = {Phys. Rev. C},
  volume = {103},
  issue = {5},
  pages = {054319},
  numpages = {12},
  year = {2021},
  month = {May},
  publisher = {American Physical Society},
  doi = {10.1103/PhysRevC.103.054319},
  url = {https://link.aps.org/doi/10.1103/PhysRevC.103.054319}
}

@article{PhysRevC.106.L021305,
  title = {Asymmetric nuclear matter and neutron star properties in relativistic ab initio theory in the full Dirac space},
  author = {Wang, Sibo and Tong, Hui and Zhao, Qiang and Wang, Chencan and Ring, Peter and Meng, Jie},
  journal = {Phys. Rev. C},
  volume = {106},
  issue = {2},
  pages = {L021305},
  numpages = {6},
  year = {2022},
  month = {Aug},
  publisher = {American Physical Society},
  doi = {10.1103/PhysRevC.106.L021305},
  url = {https://link.aps.org/doi/10.1103/PhysRevC.106.L021305}
}

@article{Sharma,
  title = {Isospin Dependence of the Spin-Orbit Force and Effective Nuclear Potentials},
  author = {Sharma, M. M. and Lalazissis, G. and K\"onig, J. and Ring, P.},
  journal = {Physical Review Letters},
  volume = {74},
  issue = {19},
  pages = {3744--3747},
  numpages = {0},
  year = {1995},
  month = {May},
  publisher = {American Physical Society},
  doi = {10.1103/PhysRevLett.74.3744},
  url = {https://link.aps.org/doi/10.1103/PhysRevLett.74.3744}
}

@article{Ebran16,
  title = {Spin-orbit interaction in relativistic nuclear structure models},
  author = {Ebran, J.-P. and Mutschler, A. and Khan, E. and Vretenar, D.},
  journal = {Phys. Rev. C},
  volume = {94},
  issue = {2},
  pages = {024304},
  numpages = {7},
  year = {2016},
  month = {Aug},
  publisher = {American Physical Society},
  doi = {10.1103/PhysRevC.94.024304},
  url = {https://link.aps.org/doi/10.1103/PhysRevC.94.024304}
}

@misc{heitz,
      title={A unified mechanism for the origin and evolution of nuclear magicity}, 
      author={L. Heitz and J. -P. Ebran and E. Khan and D. Verney},
      year={2024},
      eprint={2411.15562},
      archivePrefix={arXiv},
      primaryClass={nucl-th},
      url={https://arxiv.org/abs/2411.15562}, 
}

@article{Naito,
  title = {Comparative study on charge radii and their kinks at magic numbers},
  author = {Naito, Tomoya and Oishi, Tomohiro and Sagawa, Hiroyuki and Wang, Zhiheng},
  journal = {Phys. Rev. C},
  volume = {107},
  issue = {5},
  pages = {054307},
  numpages = {20},
  year = {2023},
  month = {May},
  publisher = {American Physical Society},
  doi = {10.1103/PhysRevC.107.054307},
  url = {https://link.aps.org/doi/10.1103/PhysRevC.107.054307}
}

@article{RadiiExp,
title = {Table of experimental nuclear ground state charge radii: An update},
journal = {Atomic Data and Nuclear Data Tables},
volume = {99},
number = {1},
pages = {69-95},
year = {2013},
issn = {0092-640X},
doi = {https://doi.org/10.1016/j.adt.2011.12.006},
url = {https://www.sciencedirect.com/science/article/pii/S0092640X12000265},
author = {Angeli, I. and Marinova, K. P.}
}

@article{Wang,
doi = {10.1088/1674-1137/abddaf},
url = {https://dx.doi.org/10.1088/1674-1137/abddaf},
year = {2021},
month = {mar},
publisher = {Chinese Physical Society and the Institute of High Energy Physics of the Chinese Academy of Sciences and the Institute of Modern Physics of the Chinese Academy of Sciences and IOP Publishing Ltd},
volume = {45},
number = {3},
pages = {030003},
author = {Wang, Meng and Huang, W.J. and Kondev, F.G. and Audi, G. and Naimi, S.},
title = {The AME 2020 atomic mass evaluation (II). Tables, graphs and references*},
journal = {Chinese Physics C}
}

@article{fissionchinoise,
  title = {Potential energy surfaces of actinide nuclei from a multidimensional constrained covariant density functional theory: Barrier heights and saddle point shapes},
  author = {Lu, Bing-Nan and Zhao, En-Guang and Zhou, Shan-Gui},
  journal = {Phys. Rev. C},
  volume = {85},
  issue = {1},
  pages = {011301},
  numpages = {5},
  year = {2012},
  month = {Jan},
  publisher = {American Physical Society},
  doi = {10.1103/PhysRevC.85.011301},
  url = {https://link.aps.org/doi/10.1103/PhysRevC.85.011301}
}

@article{grams,
	url = {https://doi.org/10.1140/epja/s10050-023-01158-6},
        doi = {10.1140/epja/s10050-023-01158-6},
	year = {2023},
	month = {nov},
	volume = {59},
	pages = {270},
	author = {Grams, Guilherme and Ryssens, Wouter and Scamps, Guillaume and Goriely, Stephane and Chamel, Nicolas},
	title = {Skyrme-Hartree-Fock-Bogoliubov mass models on a 3D mesh: III. From atomic nuclei to neutron stars},
	journal = {The European Physical Journal A}
}

@article{pillet1,
  title = {Variational multiparticle-multihole configuration mixing method applied to pairing correlations in nuclei},
  author = {Pillet, N. and Berger, J.-F. and Caurier, E.},
  journal = {Phys. Rev. C},
  volume = {78},
  issue = {2},
  pages = {024305},
  numpages = {22},
  year = {2008},
  month = {Aug},
  publisher = {American Physical Society},
  doi = {10.1103/PhysRevC.78.024305},
  url = {https://link.aps.org/doi/10.1103/PhysRevC.78.024305}
}

@article{pillet2,
  title = {Low-lying spectroscopy of a few even-even silicon isotopes investigated with the multiparticle-multihole \uppercase{G}ogny energy density functional},
  author = {Pillet, N. and Zelevinsky, V. G. and Dupuis, M. and Berger, J.-F. and Daugas, J. M.},
  journal = {Physical Review C},
  volume = {85},
  issue = {4},
  pages = {044315},
  numpages = {16},
  year = {2012},
  month = {Apr},
  publisher = {American Physical Society},
  doi = {10.1103/PhysRevC.85.044315},
  url = {https://link.aps.org/doi/10.1103/PhysRevC.85.044315}
}

@article{lebloas,
  title = {First characterization of $sd$-shell nuclei with a multiconfiguration approach},
  author = {Le Bloas, J. and Pillet, N. and Dupuis, M. and Daugas, J. M. and Robledo, L. M. and Robin, C. and Zelevinsky, V. G.},
  journal = {Physical Review C},
  volume = {89},
  issue = {1},
  pages = {011306},
  numpages = {6},
  year = {2014},
  month = {Jan},
  publisher = {American Physical Society},
  doi = {10.1103/PhysRevC.89.011306},
  url = {https://link.aps.org/doi/10.1103/PhysRevC.89.011306}
}

@article{robin1,
  title = {Description of nuclear systems with a self-consistent configuration-mixing approach: Theory, algorithm, and application to the $^{12}\text{C}$ test nucleus},
  author = {Robin, C. and Pillet, N. and Pe\~na Arteaga, D. and Berger, J.-F.},
  journal = {Physical Review C},
  volume = {93},
  issue = {2},
  pages = {024302},
  numpages = {20},
  year = {2016},
  month = {Feb},
  publisher = {American Physical Society},
  doi = {10.1103/PhysRevC.93.024302},
  url = {https://link.aps.org/doi/10.1103/PhysRevC.93.024302}
}

@article{robin2,
	title = {Description of nuclear systems with a self-consistent configuration-mixing approach. {II}. Application to structure and reactions in even-even sd-shell nuclei},
	author = {Robin, C. and Pillet, N. and Dupuis, M. and Le Bloas, J. and Pe{\~{n}}a Arteaga, D. and Berger, J.-F.},
	journal = {Physical Review C},
	volume = {95},
    number = {4},
	year = {2017},
	month = {apr},
	publisher = {American Physical Society ({APS})},
	url = {https://link.aps.org/doi/10.1103/PhysRevC.95.044315},
}

@article{robin3,
  title = {Entanglement rearrangement in self-consistent nuclear structure calculations},
  author = {Robin, Caroline and Savage, Martin J. and Pillet, Nathalie},
  journal = {Phys. Rev. C},
  volume = {103},
  issue = {3},
  pages = {034325},
  numpages = {15},
  year = {2021},
  month = {Mar},
  publisher = {American Physical Society},
  doi = {10.1103/PhysRevC.103.034325},
  url = {https://link.aps.org/doi/10.1103/PhysRevC.103.034325}
}

@article{delaroche2,
  title = {Structure of even-even nuclei using a mapped collective Hamiltonian and the \uppercase{D1S} \uppercase{G}ogny interaction},
  author = {Delaroche, J. -P. and Girod, M. and Libert, J. and Goutte, H. and Hilaire, S. and P\'eru, S. and Pillet, N. and Bertsch, G. F.},
  journal = {Physical Review C},
  volume = {81},
  issue = {1},
  pages = {014303},
  numpages = {23},
  year = {2010},
  month = {Jan},
  publisher = {American Physical Society},
  doi = {10.1103/PhysRevC.81.014303},
  url = {https://link.aps.org/doi/10.1103/PhysRevC.81.014303}
}

@article{goutte,
  title = {Microscopic approach of fission dynamics applied to fragment kinetic energy and mass distributions in $^{238}\mathrm{U}$},
  author = {Goutte, H. and Berger, J. F. and Casoli, P. and Gogny, D.},
  journal = {Phys. Rev. C},
  volume = {71},
  issue = {2},
  pages = {024316},
  numpages = {13},
  year = {2005},
  month = {Feb},
  publisher = {American Physical Society},
  doi = {10.1103/PhysRevC.71.024316},
  url = {https://link.aps.org/doi/10.1103/PhysRevC.71.024316}
}

@article{verriere2,
  title = {The Time-Dependent Generator Coordinate Method in Nuclear Physics},
  author = {Verriere, Marc and Regnier, David },
  journal = {Front. Phys.},
  volume = {8},
  pages = {233},
  year = {2020},
  month = {July},
  publisher = {American Physical Society},
  doi = {10.3389/fphy.2020.00233},
  url = {https://doi.org/10.3389/fphy.2020.00233 }
}

@article{regnier2,
  title = {From asymmetric to symmetric fission in the fermium isotopes within the time-dependent generator-coordinate-method formalism},
  author = {Regnier, D. and Dubray, N. and Schunck, N.},
  journal = {Phys. Rev. C},
  volume = {99},
  issue = {2},
  pages = {024611},
  numpages = {14},
  year = {2019},
  month = {Feb},
  publisher = {American Physical Society},
  doi = {10.1103/PhysRevC.99.024611},
  url = {https://link.aps.org/doi/10.1103/PhysRevC.99.024611}
}

@misc{Barriers,
  note = {Reference Input Parameter Library -- 2, IAEA, Vienna (2002), tecDoc (2003), unpublished; available at \url{https://www-nds.iaea.org}}
}

@misc{baldo1,
  author = {"M. Baldo},
  howpublished = {private communication},
  year = {2006}
}



\end{document}